\documentclass[12pt,preprint]{aastex}

\shorttitle{Probing Stellar Populations Using IR SBFs}
\shortauthors{Jensen et al.}

\newcommand{\kms}{km\,s$^{-1}$}
\newcommand{\kmsmpc}{km\,s$^{-1}$\,Mpc$^{-1}$}

\newcommand{\mbar}{$\overline m_{\rm F160W}$}
\newcommand{\Mbar}{$\overline M_{\rm F160W}$}
\newcommand{\vmini}{$(V{-}I)$}
\newcommand{\vminio}{$(V{-}I)_{\rm0}$}
\newcommand{\mM}{$(m{-}M)$}

\newcommand{\colorbar}{$(\overline m_I\,{-}\overline m_{\rm F160W})$}
\newcommand{\hbeta}{H${\beta}$}

\received{2002 August 14}
\begin{document}

\title{Measuring Distances and Probing the Unresolved Stellar 
Populations of Galaxies
Using Infrared Surface Brightness Fluctuations\altaffilmark{1}}
\author{Joseph B. Jensen}
\affil{Gemini Observatory,
        670 N. A`ohoku Pl., Hilo, HI  96720\\
        {\tt jjensen@gemini.edu}}
\author{John L. Tonry and Brian J. Barris}
\affil{Institute for Astronomy, University of Hawaii\\
        2680 Woodlawn Drive, Honolulu, HI  96822\\
        {\tt jt@avidya.ifa.hawaii.edu, barris@ifa.hawaii.edu}}
\author{Rodger I. Thompson}
\affil{Steward Observatory, University of Arizona, Tucson, AZ  85721\\
        {\tt rthompson@as.arizona.edu}}
\author{Michael C. Liu\altaffilmark{2}}
\affil{Institute for Astronomy, University of Hawaii\\
        2680 Woodlawn Drive, Honolulu, HI  96822\\
        {\tt mliu@ifa.hawaii.edu}}
\author{Marcia J. Rieke}
\affil{Steward Observatory, University of Arizona, Tucson, AZ  85721\\
        {\tt mrieke@as.arizona.edu}}
\author{Edward A. Ajhar}
\affil{University of Miami, Department of Physics \\
	1320 Campo Sano Drive, Coral Gables, FL  33146\\
        {\tt ajhar@physics.miami.edu}}
\and
\author{John P. Blakeslee}
\affil{Johns Hopkins University, Department of Physics and Astronomy\\
        3400 North Charles Street, Baltimore, MD  21218-2686\\
           {\tt jpb@pha.jhu.edu}}

\altaffiltext{1}{Based on observations with the NASA/ESA Hubble Space 
Telescope, obtained at the Space Telescope
Science Institute, which is operated by AURA, Inc., under NASA contract 
NAS 5-26555.}
\altaffiltext{2}{Beatrice Watson Parrent Fellow}

\clearpage

\begin{abstract}
To empirically calibrate the IR surface brightness fluctuation (SBF) 
distance scale and probe the properties of unresolved stellar populations, 
we measured fluctuations in 65 galaxies 
using NICMOS on the {\it Hubble Space Telescope}.
The early-type galaxies in this sample include elliptical and S0 
galaxies and spiral bulges in a variety of environments.
Absolute fluctuation magnitudes in the F160W (1.6 \micron) filter
(\Mbar) were derived for each galaxy
using previously-measured $I$-band SBF and Cepheid variable star distances.  
F160W SBFs can be used to measure distances to
early-type galaxies with a relative accuracy of ${\sim}$10\% provided that 
the galaxy color is known to ${\sim}0.035$ mag or better.
Near-IR fluctuations can also reveal the properties of the most
luminous stellar populations in galaxies.
Comparison of F160W fluctuation magnitudes and optical colors 
to stellar population model predictions 
suggests that bluer elliptical and S0 galaxies have significantly 
younger populations than redder ones, and may also be more metal-rich. 
There are no galaxies in this sample with fluctuation magnitudes
consistent with old, metal-poor ($t\,{>}\,$5 Gyr, [Fe/H]$\,{<}\,{-}0.7$) 
stellar population models.  
Composite stellar population models imply that bright fluctuations in the 
bluer galaxies may be the result of an episode of recent star formation 
in a fraction of the total mass of a galaxy.
Age estimates from the F160W fluctuation magnitudes are consistent with
those measured using the \hbeta\ Balmer line index.  
The two types of measurements make use of completely different techniques 
and are sensitive to stars in different evolutionary
phases.  Both techniques reveal the presence of intermediate-age 
stars in the early-type galaxies of this sample.

\end{abstract}


\section{Introduction}
The techniques for measuring surface brightness fluctuations (SBFs)
were developed primarily with the goal of determining extragalactic
distances.  The optical SBF method has proven to be remarkably
useful for estimating distances (see Blakeslee et al. 1999 for a review).
The initial motivation for extending ground-based SBF 
measurements to near-IR wavelengths
was to take advantage of intrinsically brighter fluctuations and
better atmospheric seeing to reach much greater distances
(Jensen et al. 2001, hereafter J2001; Jensen, Tonry, \& Luppino 1999; 
Liu \& Graham 2001).  
With {\it Hubble Space Telescope's} (HST) Near Infrared Camera 
and Multi-Object Spectrometer (NICMOS), 
the low background and lack of atmospheric seeing
offer significant advantages over ground-based IR SBF measurements.

Accurate SBF distance measurements rely on the empirical calibration
of absolute fluctuation amplitudes.  
Several studies have used observations of galaxies 
with previously-measured distances to determine absolute fluctuation 
magnitudes, which were found to agree with the 
predictions of stellar population models.
The first IR SBF measurements were made by Luppino \& Tonry (1993),
who found that M32 has a somewhat brighter (${\sim}0.25$~mag) 
$K$-band SBF magnitude than the bulge of M31.
Pahre \& Mould (1994) measured $K$-band SBF magnitudes for a sample of Virgo
galaxies, and found that two of eight had significantly brighter
(${\gtrsim}1$~mag) fluctuations than the others.  
Pahre \& Mould's results were confirmed by Jensen, Luppino, \& Tonry (1996)
in their sample of seven Virgo cluster elliptical galaxies.
The IR SBF signal for one of the two anomalous galaxies (NGC~4365) was
later found to have been over-estimated due to the contribution
of undetected globular clusters (Jensen, Tonry, \& Luppino 1998); 
the other (NGC~4489) still appeared to have brighter fluctuations
than the others.

To the 1996 Virgo sample, Jensen et al. (1998) added galaxies in the Fornax
and Eridanus clusters.  
Three of the bluest galaxies in their sample showed $K$-band fluctuation 
magnitudes that were ${\sim}0.25$~mag brighter than the others.
Jensen et al. noted that the models implied younger stellar populations
in these bluer ellipticals. 
However, the sample size was small and the range in \vmini\ color was limited;
hence, the slope they measured was not statistically significant and they
adopted a constant $K$-band SBF calibration.
The $K$-band fluctuations for the bluest galaxy in their sample (NGC~4489)
were significantly brighter than the others in the Virgo cluster, but
the measurement was not trusted due to its low signal-to-noise ($S/N$) ratio.
Mei, Silva, \& Quinn (2001c) subsequently re-observed this galaxy
and found comparably bright $K$-band fluctuations.

Recently, a larger sample of $K$-band SBF magnitudes for Fornax
cluster galaxies has been measured by Liu, Graham, \& Charlot
(2002).  Combined with the previously published data, the resulting sample
covered a much larger range of galaxy properties (e.g., color
and luminosity).  Liu et al. (2002) discovered that 
the brighter $K$-band fluctuations seen in bluer galaxies
in their sample and in earlier data sets
(Luppino \& Tonry 1993; Pahre \& Mould 1994; Jensen et al. 1998)
were correlated with \vmini.
Liu et al. found the $K$-band slope with color to be comparable to
that for $I$-band SBFs.  
This relation naturally explained the bright IR SBFs previously 
measured in M32 (Luppino \& Tonry 1993) and NGC~4489 (Mei et al. 2001c). 
Moreover, a generic prediction of IR SBF models is that IR fluctuations
are sensitive to variations in the ages and metallicities of stellar 
populations.
Hence, the discovery of a {\it systematic} relation between IR SBFs and 
galaxy color means that age and metallicity are related in a way
that reflects the star formation history of a galaxy.
Liu et al. (2002) concluded that early-type galaxies in clusters
have a significant spread in age and approximately solar metallicities.
  
Liu et al. (2002) also reported finding three 
Fornax cluster galaxies with $K$-band fluctuation magnitudes brighter than 
the other galaxies with similar \vmini\ colors.  These galaxies were not 
included in their calibration fit.  
They are primarily lower luminosity galaxies, and models suggest the 
presence of a high-metallicity burst of star formation in the last few Gyr.  
Mei et al. (2001a) have confirmed that one of the three Fornax galaxies 
(NGC~1427) has unusually bright $K$-band fluctuations.

To calibrate the F160W (1.6 \micron) SBF distance scale and 
better understand the nature of the bluer, low-luminosity elliptical
galaxies, we measured F160W SBF magnitudes in a large sample of 65 galaxies
spanning a wide range in color (\vminio$\,{=}\,1.05$ to 1.28).
All the data presented here were obtained using the NIC2 camera
of NICMOS. 
The data are of uniform image quality, and the 
$S/N$ ratios are large.  Many of the galaxy images were collected for other
programs and retrieved from the public archive.

Distances to the galaxies in this sample were taken from the 
growing collection of Cepheid variable star distances measured with the
$HST$ (Freedman et al. 2001, Ferrarese et al. 2000b; 
Gibson \& Stetson 2001; Saha et al. 2001), 
and from the extensive $I$-band SBF survey (Tonry et al. 2001; 
Ajhar et al. 1997; Lauer et al. 1998; Tonry et al. 1997). 
The $I$-band SBF distances used here were calibrated using
Cepheid distances to the same galaxies; no assumptions about
group or cluster membership were made (Tonry et al. 2001).
For the current study we shifted the $I$-SBF distances 
published by Tonry et al. by $-0.16$~mag to the
new Cepheid zero point of Freedman et al. (2001), which
makes use of the improved Cepheid period-luminosity relations 
published by Udalski et al. (1999).
The distances presented in this paper, whether from Cepheids directly 
or from $I$-band SBFs, are subject to the systematic
uncertainties in the Cepheid distance scale (Ferrarese et al. 2000a).
We adopted a distance modulus to the Large Magellanic Cloud (LMC) of 
18.50 mag (Freedman et al. 2001; Carretta et al. 2000), and acknowledge
that the continuing debate over the LMC distance remains one of the
largest sources of systematic uncertainty.

In this paper we present a new calibration of the F160W distance
scale and explore its sensitivity to galaxy color. 
We also compare the absolute fluctuation magnitudes to those
predicted by three sets of stellar population models:
the widely-used models of Worthey (1994), 
the Bruzual \& Charlot (1993, 2002) models (as published by 
Liu, Charlot, \& Graham 2000 and Liu et al. 2002),
and the Vazdekis (1999, 2001) models (as published by Blakeslee,
Vazdekis, \& Ajhar 2001).
SBF ages determined using stellar population models are
compared with those measured using the age-sensitive \hbeta\ Balmer
line strength.

\section{Observations}
\subsection{F160W SBF Measurements \label{observingsec}}

The observations of galaxies used in this study come from two basic
types of NICMOS program, each of which accounts for roughly half the
data.  The first set of observations came from two programs explicitly
designed for SBF measurements.  The first of these (NICMOS program ID
7453) targeted galaxies in Virgo, Leo, and Fornax to empirically
calibrate the F160W SBF distance scale for more distant measurements.
The second (program ID 7458) measured fluctuations in a large sample
of Fornax galaxies at multiple wavelengths to better understand 
stellar populations.  A subset of the measurements presented in this
paper was published by J2001.
The second category of
observations is comprised of data taken from the public archive from
a variety of programs.  
Most of these are short, ``snapshot'' survey images of the centers of 
galaxies, and have only a minimal number of individual exposures.
Observational data are listed in Table~\ref{observationtable}.

The observations presented in this study were taken with the NIC2
camera (19.2 arcsec field of view) through the F160W filter.  
F160W NIC2 observations of approximately 300 galaxies with 
heliocentric redshifts 
less than 10,000 \kms\ were retrieved from the public archive and examined.  
A smooth fit to each galaxy was subtracted 
from the pipeline-reduced image and the residual frame examined.  
All the galaxies that were
judged to be hopelessly dusty were rejected from further consideration.
Data for the remaining ${\sim}200$ galaxies were reduced again from 
the raw data and the SBF analysis completed following the same procedures
as described by J2001.  
The subset of 65 galaxies presented here are those for which reliable 
distances are known from either Cepheid variable stars or $I$-band SBFs.

The methodology for determining F160W fluctuation magnitudes was very
similar to that described by J2001, with some simplifications.
We used the software developed by the NICMOS GTO team
to prepare the images for analysis (Thompson et al. 1999).  
Dark current was first subtracted from the raw NIC2 F160W images.  
The multiple reads of each MULTIACCUM sequence
were combined and cosmic rays identified and removed.
Adjustment of the bias pedestal for each quadrant was performed
as described by J2001, although it was unnecessary in most cases.
Exposures from individual telescope pointings were flat field
corrected and combined.
Residual images from cosmic rays were not a significant source
of contamination in these high-$S/N$ ratio data.

The photometric zero point used by J2001 was determined by 
M. Rieke and the NICMOS team.
Additional standard star measurements combined with a better
tie to ground-based photometry has yielded a better zero point
for the NIC2 F160W filter.  
The new photometric zero point that we adopted for this study
is 0.033 mag fainter than that used by J2001.  
The new calibration for NIC2 is 
$2.126{\times}10^{-6}$ Jy\,ADU$^{-1}$\,s$^{-1}$. 
The magnitude zero point on the Vega system is 1083 Jy.
This new calibration is within the uncertainty in zero point
published by J2001.
If the current zero point were applied to the J2001 data without
any other changes to the calibration, the resulting
fluctuation magnitudes would be 0.03 mag fainter, and the
Hubble constant would be larger by 1.6\%.
SBF magnitudes were corrected for Galactic extinction using 
the measurements of Schlegel, Finkbeiner, \& Davis (1998). 
We adopted $A_B\,{=}\,4.315\,E(B{-}V)$ and $A_H = 0.132 A_B$ 
(Schlegel et al 1998). 

The galaxies in this sample are much closer than the distant
galaxies used by J2001 to determine the Hubble constant 
(5 included here were taken from the intermediate-distance sample of J2001). 
The median $S/N$ ratio was 16 per pixel for this data set,
which, given the large number of pixels sampled, is more than
sufficient to achieve a highly-reliable measurement. 
Most of the data had $S/N$ ratios between 10 and 20; the full
range includes measurements with $S/N$ ratios as low as 5 and 
as high as 4000.
The fluctuation power was determined by fitting the scaled
power spectrum of the reference point-spread function (PSF) to
the spatial power spectrum of the cleaned and galaxy-subtracted data.
The library of empirical PSF stars collected 
as part of the IR SBF Hubble constant project (J2001)
was used to perform the SBF analysis.  A full discussion of the
techniques for calculating fluctuation magnitudes \mbar\ from
NICMOS data is presented by J2001.

We found that correcting the F160W SBF magnitudes for undetected globular 
clusters or background galaxies was unnecessary for the relatively-nearby
galaxies of this study because the stellar SBF signal
always dominated over other sources of variance.  
The final galaxy-subtracted images were masked of visible point sources and 
dusty regions before proceeding with the SBF analysis.  
Only the five intermediate-distance galaxies from
J2001 required corrections for undetected globular clusters, 
background galaxies, and residual background patterns (as
described by J2001). 

The uncertainties in apparent fluctuation magnitudes \mbar\ were typically
0.1 mag or less.  The primary components of the uncertainty were 
the fit to fluctuation power and the PSF normalization 
(usually 0.05 mag each).  
The contribution from sky subtraction was much smaller (typically 0.01 mag).
Contributions to the uncertainty from undetected globular clusters or 
background galaxies were negligible (${<}0.01$~mag), 
and contaminating power from residual cosmic rays or incomplete bias 
subtraction were unmeasurable in these high-$S/N$ ratio measurements.

\subsection{Distances and Absolute Fluctuation Magnitudes \label{distancesec}}

Computing the absolute fluctuation magnitudes \Mbar\ for the 
galaxies in our sample required independent distance measurements.
All the distance moduli used in this study (Table~\ref{mbartable}) 
were based, either directly or indirectly, on Cepheid variable 
star distances.  
Two classes of distance measurements are presented here:
first, a large set of 61 $I$-band SBF distances (calibrated using
Cepheid distances as described by Tonry et al. 1997),
and second, a smaller set of nine Cepheid distances measured 
using WFPC2 on the {\it HST}.

Because only a few of the spiral galaxies with measured Cepheid distances
have smooth, dust-free regions appropriate for SBF analysis, the majority 
of the absolute fluctuation magnitudes presented in this paper 
were computed using $I$-band SBF distances (Tonry et al. 2001).  
Most of the $I$-band SBF distance measurements were
made using ground-based telescopes, but five galaxies were observed
using WFPC2 (J2001; Lauer et al. 1998; Ajhar et al. 1997).  
The empirical $I$-band SBF calibration 
(Tonry et al. 1997) adopted for this study used Cepheid and $I$-SBF 
distance measurements to 
seven galaxies for which both types of measurement are possible.  
No assumptions about group or cluster membership were made
to connect the Cepheid and $I$-band SBF distance scales.
The original Key Project calibration of the Cepheid distance scale 
(Ferrarese et al. 2000b, Freedman \& Madore 1990) was 
used for the empirical $I$-band SBF calibration presented by 
Tonry et al. (1997, 2001).
For the $I$-band SBF distances presented here we have updated the 
zero point using the greatly-improved period-luminosity relations 
determined by Udalski et al. (1999) using 650 LMC Cepheids.  
The new period-luminosity relation results in a shift of $-0.16$ mag
in the distance moduli of all the $I$-band SBF galaxies.
No metallicity correction to the Cepheid distances is adopted for 
this study.

The modifications to the period--luminosity relation adopted by
Freedman et al. (2001) are distance dependent.  
The new period--luminosity relation is different for the $I$-band, 
while at $V$ it is unchanged (Udalski et al. 1999).  The effect
is that the new period--luminosity relation predicts higher reddening 
corrections for redder Cepheids, and therefore smaller distances.
The effect is largest in Cepheids with longer periods.  Because only
the brightest (longest-period) Cepheids are detected in the most
distant galaxies, the offset between the previously-published distances
and those derived using the new calibration are largest in the most
distant galaxies (${\sim}0.2$ mag, or 10\% in distance).
The offset to the $I$-band SBF calibration due to the new
period-luminosity relation is 0.16 mag, or 8\% in distance.

Nine galaxies in the F160W NICMOS sample have
Cepheid distances measured using HST (Ferrarese et al. 2000b; 
Freedman et al. 1994, 2001; Gibson \& Stetson 2001; 
Gibson et al. 1999, 2000; Graham et al. 1997; Saha et al. 1996, 2001).  
A distance modulus to
the LMC of 18.50 mag was adopted for the
current study and for the Key Project papers (Ferrarese et al. 2000b;
Freedman et al. 2001).  This distance to the LMC appears justified 
in light of many recent distance measurements (e.g.,  Carretta et al. 2000), 
although significant differences certainly remain between techniques 
and investigators (Udalski et al. 1999; others). 
A full discussion of the LMC distance issue is beyond the scope 
of this paper;
future adjustments to the LMC distance modulus of 18.50 should be 
applied as a constant offset to all the distances used herein
and to the resulting F160W SBF calibration.
The conclusions of this paper are not affected by an uncertainty
of 0.15~mag in the LMC distance.

Freedman et al. (2001) also applied an empirical metallicity correction 
of ${-}0.2\,{\pm}\,0.2$ mag~dex$^{-1}$ in metallicity
to the Cepheid calibration such that more metal-rich Cepheids are
intrinsically brighter
(Freedman et al. 2001; Gibson \& Stetson 2001; Kennicutt et al. 1998).  
For comparison, separate distances for the Key Project galaxies are 
listed in Table~\ref{mbartable} using the new period--luminosity relation
alone (``new PL'', Freedman et al. 2001; Gibson \& Stetson 2001), 
and using both the new period--luminosity relation and empirical 
metallicity correction (``new PL+Z'', Freedman et al. 2001).  
We chose not to adopt the empirical metallicity correction to the
Cepheid distances endorsed by Freedman et al. (2001) because it is
not yet entirely clear that the metallicity correction is justified.
Udalski et al. (2001) find no evidence for a trend in the luminosity
of Cepheids with metallicity.  Another study based on theoretical
models of Cepheid structure predicts a correction to the period--luminosity
relation of ${+}0.27$ mag~dex$^{-1}$ in metallicity, which is similar in 
magnitude but {\it opposite} in sign from the empirical relation
adopted by Freedman et al. (Caputo, Marconi, \& Musella 2002).  
The apparently better agreement with
the reliable maser distance to the galaxy NGC 4258 is offered as 
evidence that the theoretical relationship is more realistic. 
Of course a slightly smaller distance modulus to the LMC would
also explain the difference, so the maser distance to NGC~4258 cannot
be regarded as evidence that the metallicity correction must have a
particular sign or magnitude.
Given the uncertainty in the metallicity correction at this point,
we chose not to adjust the $I$-band SBF calibration.
In the end, the size of the metallicity 
correction is not very important provided that the correction is
applied consistently through the LMC rung of the distance ladder.
We note that the metallicity-corrected Cepheid distances
reported by Freedman et al. (2001) are based on Cepheid measurements of
LMC Cepheids that have not been corrected for metallicity.  Given the 
similarity in metallicity between Galactic Cepheids and most
of the spiral galaxies in which Cepheids have been detected, the 
effect of a fully-consistent metallicity-corrected Cepheid distance
scale and the uncorrected calibration we adopted for this paper is minimal
(approximately 0.02~mag).  
The data plotted in the figures are based on distances that 
include no metallicity corrections.

The uncertainties in \mbar\ were combined in quadrature with the 
uncertainties in distance modulus to get the final uncertainties 
in \Mbar\ presented in Table~\ref{mbartable}.
The median uncertainty in \Mbar\ was 0.18 mag for this sample.
The estimated systematic error in the Cepheid distance
scale, which we incur regardless of which distances are used to 
compute \Mbar, is 0.16 mag (Ferrarese et al. 2000a), 
but could be larger due to blending or other effects 
(Gibson \& Stetson 2001; Mochejska et al. 2000; Ferrarese et al. 2000c).
The dominant sources of systematic uncertainty are the distance
to the LMC (0.13 mag) and the WFPC2 photometric zero point (0.09 mag).
The systematic uncertainty is {\it not} included in the uncertainty 
in \Mbar\ listed in Table~\ref{mbartable}.
It is important to note that the uncertainty in \Mbar\ is correlated
with the uncertainty in \vmini\ for the values derived from
$I$-band SBF distances.  \Mbar\ is a function of the $I$-SBF distance 
modulus, which is a function of \vmini\ color (Tonry et al. 1997).

\subsection{Galaxy \vmini\ Colors}

To calibrate the F160W SBF distance scale, we used the optical \vminio\
color corrected for extinction to constrain the fluctuation magnitude
dependence on stellar population.  
Most of the \vminio\ color data in Table~\ref{mbartable} 
were taken from the optical $I$-band SBF survey (Tonry et al. 2001).  
Galactic extinction corrections were made using the 
Schlegel et al. (1998) extinction maps.
The data were collected using ground-based telescopes in annular
regions that were typically much larger than the central 20-arcsec 
regions imaged using NIC2.  
In a few cases we checked the optical colors within the NIC2 field of 
view to ensure that color gradients within individual galaxies 
were not significant.  
For the early-type galaxies in this sample, color gradients  
do not produce significant color differences
between the F160W and $I$-band regions.  A few galaxies, 
including the spirals not included in the $I$-band SBF survey 
(Tonry et al. 2001), were re-observed to allow 
a direct \vminio\ color measurement within the NIC2 field of view.

\section{Using F160W SBF Magnitudes to Measure Distances \label{calibsec}}

The primary motivation for exploring the variation of \Mbar\ with
stellar population was to better calibrate F160W SBFs as a distance 
indicator.
With the high spatial resolution and low background achieved using NICMOS,
J2001 demonstrated that fluctuations could be measured
in modest exposures (one or two orbits) to distances beyond 100 Mpc.
J2001 utilized a limited subset of the current data set
and the Cepheid zero point calibration of Ferrarese et al. (2000b)
to empirically calibrate the F160W SBF distance scale.
They measured distances to a sample of galaxies reaching redshifts 
of 10,000 \kms\ for the purpose of determining the Hubble constant.  
J2001 restricted their calibration to galaxies redder than 
\vminio$\,{>}\,1.16$ and found no significant slope of \Mbar\ with galaxy
color.  

\begin{figure}
\plotone{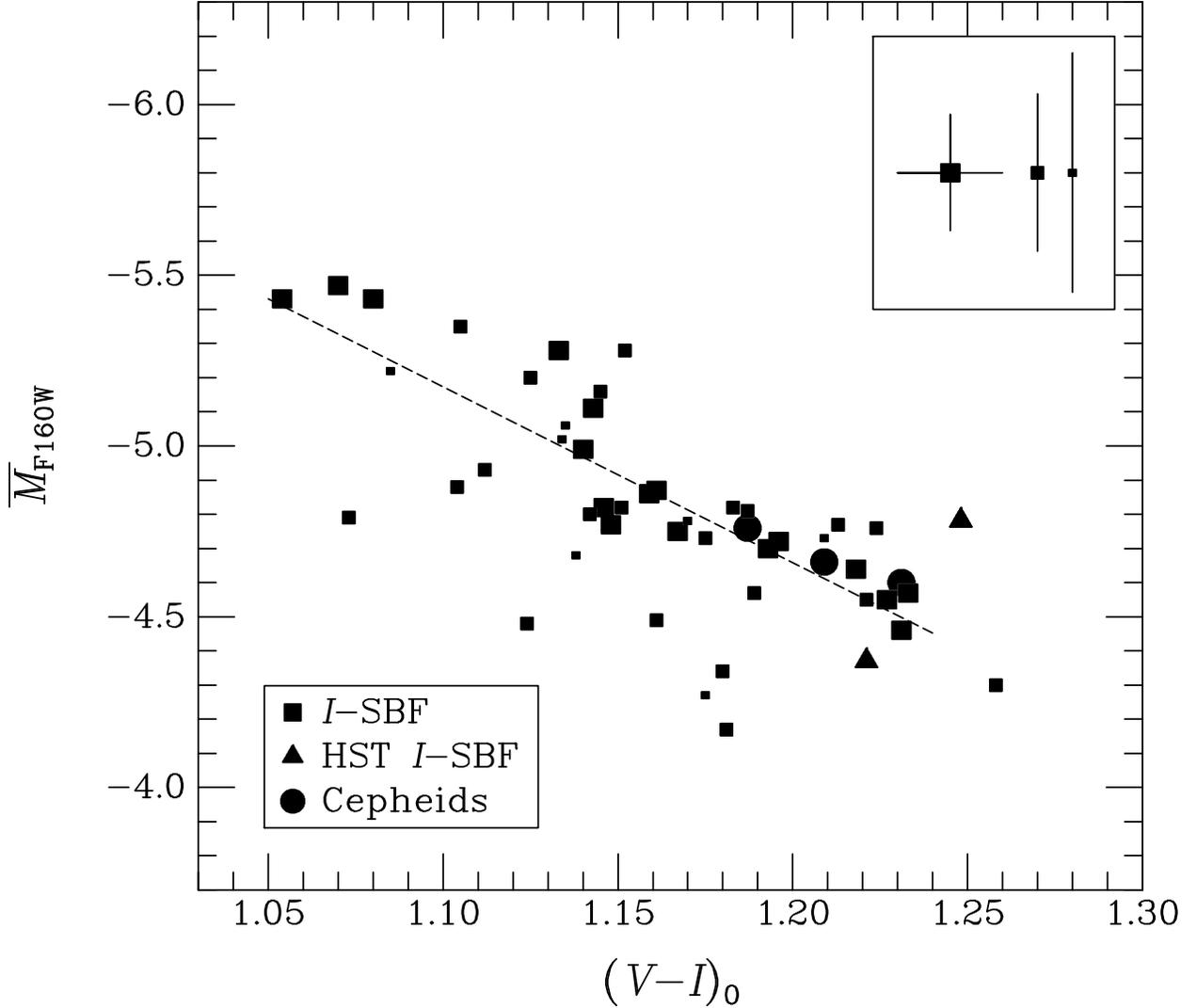}
\figcaption[]{Absolute fluctuation magnitudes \Mbar\ plotted vs. 
the extinction-corrected \vminio\ color for the 47 galaxies that
show no signs of dust in the NIC2 field of view.  
The square points represent \Mbar\ values derived using ground-based
$I$-band SBF distances, and the triangles indicate galaxies 
with $I$-band SBF distances 
measured with WFPC2 on the HST.  
The size of each point indicates the uncertainty in \Mbar.  
The largest points have uncertainties less than 0.2 mag, the medium-sized
points fall between 0.2 and 0.3 mag, and the smallest points have
uncertainties greater than 0.3 mag.  Median error bars for each point
size are shown at the top of the figure.
The circles indicate three galaxies with reliable Cepheid distances
(NGC~224=M31, NGC~3031=M81, and NGC~4725).
They are also plotted using their $I$-band SBF distances.
\label{calibfig}}
\end{figure}

A constant \Mbar\ calibration is inappropriate for the full
color range spanned by the galaxies in this sample (Fig.~\ref{calibfig}).
For the full data set, we fitted a slope adopting the maximum likelihood
method described by Liu et al. (2002) to account for the correlated
uncertainties in \Mbar\ and \vmini.  
Uncertainties are correlated because \Mbar\ depends on $I$-band
SBF distances, which are in turn computed using the galaxy \vmini\
color.  The Liu et al. (2002) method accounts for this non-zero
covariance between \Mbar\ and \vmini\ in determining the best-fitting
slope and intercept.  If the covariance were ignored, the slope of the 
fitted line would be biased (see Liu et al. 2002 for a discussion).  

We chose to restrict the sample of galaxies used for calibration
purposes to those that show no sign of dust in the NIC2 
field of view because clumpy dust makes a galaxy look bumpier, and 
hence fluctuation magnitudes brighter.
Many of the rejected galaxies have well-defined dust
lanes that can be masked.  
The SBF magnitudes are most likely unaffected by the dust, but we 
exclude them from the calibration fit to be safe.  
Given the large size of the sample, we
can afford to be conservative in excluding dusty galaxies for the
purpose of computing the calibration.  
Fitting the 47 dust-free galaxies gives
\begin{equation}
\overline M_{\rm F160W} = (-4.86\pm0.03) + (5.1\pm0.5)[(V{-}I)_0\,{-}\,1.16]
\ \ {\rm for}\ \ 1.05\,{<}\,(V{-}I)_0\,{<}\,1.24.
\label{calibeq}
\end{equation}
The ${\chi}^2$ per degree of freedom is 1.17, indicating that the 
uncertainties are reasonable.  
The slope of $5.1\,{\pm}\,0.5$ mag per mag in \vmini\
color is steep, and similar to the slope of $4.5\,{\pm}\,0.25$
measured at $I$-band (Tonry et al. 1997).  Liu et al. (2002)
found a slope of $3.6\,{\pm}\,0.8$ at $K$ (2.2 \micron). 
Stellar population models initially predicted that the
slope of IR fluctuation magnitude with color would be opposite that
at $I$-band (Worthey 1994), but it is now clear that a significant
slope of the {\it same} sign persists to 2.5 \micron.  The implications
of a positive slope are discussed in Section~\ref{modelsec}.

The relation in Equation~\ref{calibeq} is based on $I$-band SBF 
distances calibrated
using the improved Cepheid period-luminosity relation of
Udalski et al. (1999).  
The Cepheid distances from Freedman et al. (2001) without metallicity
corrections were adopted, and we used the same LMC distance modulus 
of 18.50 mag as Freedman et al.
In Figure~\ref{calibfig} symbols of three sizes are plotted, 
with the largest symbols for galaxies with uncertainties less than 
0.2 mag and the smallest for measurements with uncertainties greater 
than 0.3 mag.  Median uncertainties in each bin are plotted.
Three galaxies in the ``dust-free'' sub-sample also have Cepheid
variable star distances.  They were not included in the fit
because they are not independent of the $I$-band SBF
measurements, and their uncertainties arise from different sources.
They are, however,
plotted in Figure~\ref{calibfig} to show that the calibration
derived using $I$-band SBF distances agrees with direct Cepheid distance
measurements.

The value of \vminio$\,{=}\,1.16$ marks the color at 
which the strong slope observed in the bluer
galaxies makes way to an apparent flattening at the red end.
Fits restricted to galaxies with \vminio$\,{>}\,1.16$ are 
consistent with a slope of zero, as found by J2001
using a smaller calibration sample.  
The sloping fit over the entire color range spanned by the sample, 
however, is statistically robust.
The brightest cluster galaxies observed by J2001 are all redder than 
\vminio$\,{=}\,1.16$, and the use of a constant \Mbar\ 
by J2001 was justified.
The stellar population models presented in the following sections
suggest that the steep slope shown in Figure~\ref{calibfig}
cannot continue to arbitrarily red values of \vmini because galaxies
cannot be made of arbitrarily old stars.
We therefore adopt the relation
between \Mbar\ and \vminio\ given in equation \ref{calibeq} and 
emphasize that it is {\it only} applicable to galaxies with \vmini\ colors
between 1.05 and 1.24.

The small statistical uncertainty in the zero point and the value
of ${\chi}^2$ close to 1 suggests that the ``cosmic scatter'', or
the variation in \Mbar\ not accounted for with the single \vmini\
parameter, is small.  The calibration presented here suggests that
relative distances to galaxies with \vmini\ colors typical of elliptical
galaxies can be measured with 10\% accuracy or better provided that
the uncertainty in \vminio\ is 0.035~mag or smaller, 
and that the photometric zero point is known to 0.05~mag or better.  
This estimate of the statistical uncertainty
does not include the systematic uncertainty in the distance scale
zero point, which includes the uncertainties in the Cepheid distance
scale and the distance to the LMC.
The scatter increases significantly when dusty galaxies are included.
The scatter below the best-fit line in Figure~\ref{calibfig} 
appears larger than above it.  This
could be caused by the F160W SBF amplitude being under-estimated, or the
$I$-band SBF amplitude being over-estimated (and the distances 
under-estimated).  The uncertainties in the $I$-band SBF magnitudes are 
larger than 0.2 mag for most of the outlying points, but there is no 
obvious sign of a systematic problem with the $I$-band SBF measurements.  
The fit has been appropriately weighted by the uncertainties, and is
consistent with the most accurate measurements.  Excluding them would
not significantly change the relation in Equation~\ref{calibeq}.

J2001 used a constant fluctuation magnitude \Mbar$\,{=}\,{-}4.86$
in their determination of the Hubble constant.  It is fair to ask
what effect the new sloping calibration and revised Cepheid
calibration would have if they had been adopted instead.  
To address this issue,
we selected the five galaxies from the J2001 sample that fall
within the valid color range \vminio$\,{<}\,1.24$ of the current distance 
calibration and calculated the Hubble constant for the preferred
flow models using exactly the same techniques used by J2001.
The increase in the Hubble constant that results from the changes
in the zero point and slope of the calibration is not sensitive to the
exact flow model adopted or rejection of low-$S/N$ ratio observations
(see J2001 for a discussion of these effects).
Because the new Cepheid calibration and NIC2 zero point result
in a significantly fainter calibration zero point, the value of the Hubble 
constant would increase by 10\% from 76 or 77 to 85 \kmsmpc. 
If the new calibration is applied to only the most distant 
galaxies from J2001, the Hubble constant would increase from 
72~\kmsmpc\ to 79.  
Only a small fraction (1.5\% of the 10\%) of the change is due to the 
slope of the calibration;
the 0.16 mag change in the Cepheid zero point resulting from the
adoption of the improved period-luminosity relation is the dominant factor.
If we had adopted the metallicity corrections endorsed by Freedman et al.
(2001), the increase in the Hubble constant would have been only 5\%
(corresponding to a zero point shift of 0.06 mag instead of 0.16 mag).
All the changes described happen to 
affect the calibration in the same direction, i.e., reducing the
absolute brightness of the fluctuations and decreasing the implied
distances.


\section{Stellar Population Models \label{modelsec}}

Infrared SBF magnitudes provide important new constraints on
theoretical and semi-empirical stellar population models.  
The models provide insight into why fluctuation magnitudes
vary with color the way they do.
Each single-burst, constant-metallicity model
is constructed by adopting an initial mass function and uniform
composition for an ensemble of theoretical stars.  These stars are
then allowed to evolve according to the constraints and assumptions
of the particular model.  At each time step, the luminosity function
of the ensemble of stars is integrated to determine observable values,
including broad-band colors, line indices, and fluctuation magnitudes.  
Luminosity fluctuations
in a galaxy are proportional to the second moment of the luminosity
function (Tonry \& Schneider, 1988), and can be computed from
the model luminosity function at each time step.  Because fluctuations
are weighted to the most luminous stars in the population,
SBFs provide a way to better measure the contributions of luminous
first-ascent red giants and asymptotic giant branch (AGB) stars in 
unresolved stellar populations.

The first detailed models to successfully predict SBF magnitudes were 
those of Worthey (1994).
SBF predictions from the models of Bruzual \& Charlot (Liu et al. 2000, 2002)
and Vazdekis (Blakeslee et al. 2001) have recently been 
published.
In this paper, we compare empirical F160W fluctuation magnitudes 
\Mbar\ to the three sets of models to learn more about 
the relative ages and metallicities of the galaxies in our sample.  
These three sets of models are the most successful models in 
common use today for which SBF magnitudes have been published.
We also compare the model predictions to the measured fluctuation 
color \colorbar, which is independent of the distance measurements.
Agreement between observed fluctuation magnitudes and stellar population
model predictions can therefore be used to simultaneously confirm
the reliability of SBF distance measurements and reinforce the age
and metallicity interpretations.

\subsection{The Bruzual \& Charlot Models}

\begin{figure}
\plotone{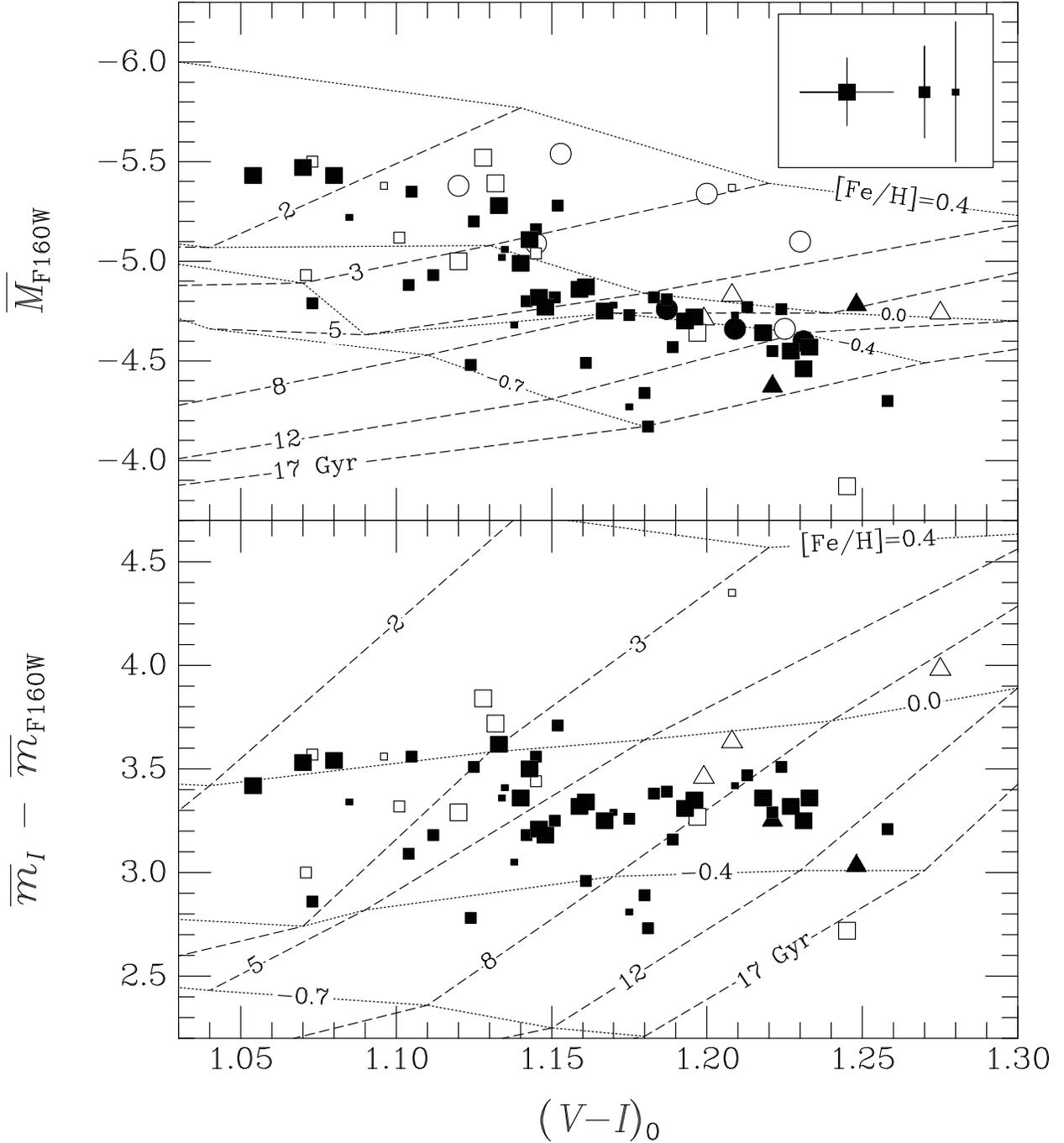}
\figcaption[]{
The upper panel shows absolute fluctuation magnitudes plotted vs. 
galaxy optical color.  The circular points were
calculated using Cepheid distances, the square points using ground-based
$I$-band SBF distances, and the triangles using HST $I$-band SBF distances. 
Open symbols indicate galaxies that show signs of dust.
Clumpy dust can cause fluctuations to appear brighter.
The lines behind the points 
indicate Liu et al. (2000, 2002) models of constant age 
(dashed lines) and metallicity (dotted lines). 
The ages (Gyr) and metallicities ([Fe/H]) of the models are indicated. 
The lower panel shows the distance-independent fluctuation color
as a function of \vminio.
\label{liufig}}
\end{figure}

Absolute F160W fluctuation magnitudes are plotted as a function of
extinction-corrected \vminio\ color in the top panel of Figure~\ref{liufig}.
Data for all 65 galaxies are plotted, including galaxies containing dust
that were excluded from the calibration fits.  Points for galaxies with
Cepheid distances are included as well.  
In the lower panel of Figure~\ref{liufig}
we plotted the fluctuation color \colorbar\ as a function of galaxy
\vmini\ color.  The fluctuation color is independent of distance, and
therefore provides additional insight into the stellar populations of
the galaxies without being subject to the systematic uncertainties of
the distance measurements.

Theoretical fluctuation magnitudes and colors from Liu et al. (2000, 2002)
are shown in Figure~\ref{liufig} as dashed (constant age) and dotted
(constant metallicity) lines. 
The Liu et al. (2002) SBF predictions were based on the single-burst 
population synthesis models of Bruzual \& Charlot (1993, 2002), originally 
described by Liu et al. (2000), and then slightly improved by 
Liu et al. (2002).  
The models provide several choices for the input evolutionary tracks 
and spectral libraries; a detailed comparison of the various options is 
given by Liu et al. (2000).  
We use their preferred set of inputs: the Padova evolutionary tracks of
Bertelli et al. (1994), semi-empirical spectral energy distributions
from Lejeune, Cuisinier, \& Buser (1997), and a Salpeter initial mass 
function.

The Bruzual \& Charlot models (Liu et al. 2002) 
imply that the trend towards brighter 
fluctuations in bluer galaxies is consistent
with younger stellar population models with high metallicities.  The
redder galaxies appear older and more metal-poor than the bluer ones
(model ages are relative to the age of the Universe,
thus a model age of 17 Gyr does not imply that a population formed
before the Universe was born).
A similar trend has been seen in $K$-band fluctuations (Jensen et al.
1998; Mei, Quinn, \& Silva 2001b; Liu et al. 2002).
It is interesting to note that there are no fluctuation magnitudes 
among the bluer galaxies that are consistent with stellar population models 
older than about 5 Gyr.
None of the galaxies appear to have metallicities lower than 
approximately [Fe/H]$\,{=}\,{-}0.7$.
The spread in fluctuation magnitudes with color traces a line of
models with nearly constant metallicity down to \vminio$\,{=}\,1.15$;
bluer galaxies appear more metal-rich than those with \vminio$\,{>}\,1.15$.  

The distance-independent plot of fluctuation
colors in the lower panel of Figure~\ref{liufig} 
leads to the same conclusion about the
relative ages of galaxies with different colors.  
The absolute metallicities implied by the \colorbar\ models are 
very similar to those in the upper panel.
The bluer galaxies have the same fluctuation magnitudes as
stellar population models with significantly younger ages and slightly 
higher metallicities than redder galaxies.
Agreement between the \Mbar\ and the distance-independent 
\colorbar\ models suggests that the total systematic error in the 
distance scale calibration is likely to be of order 0.1~mag.


\subsection{The Vazdekis Models}

\begin{figure}
\plotone{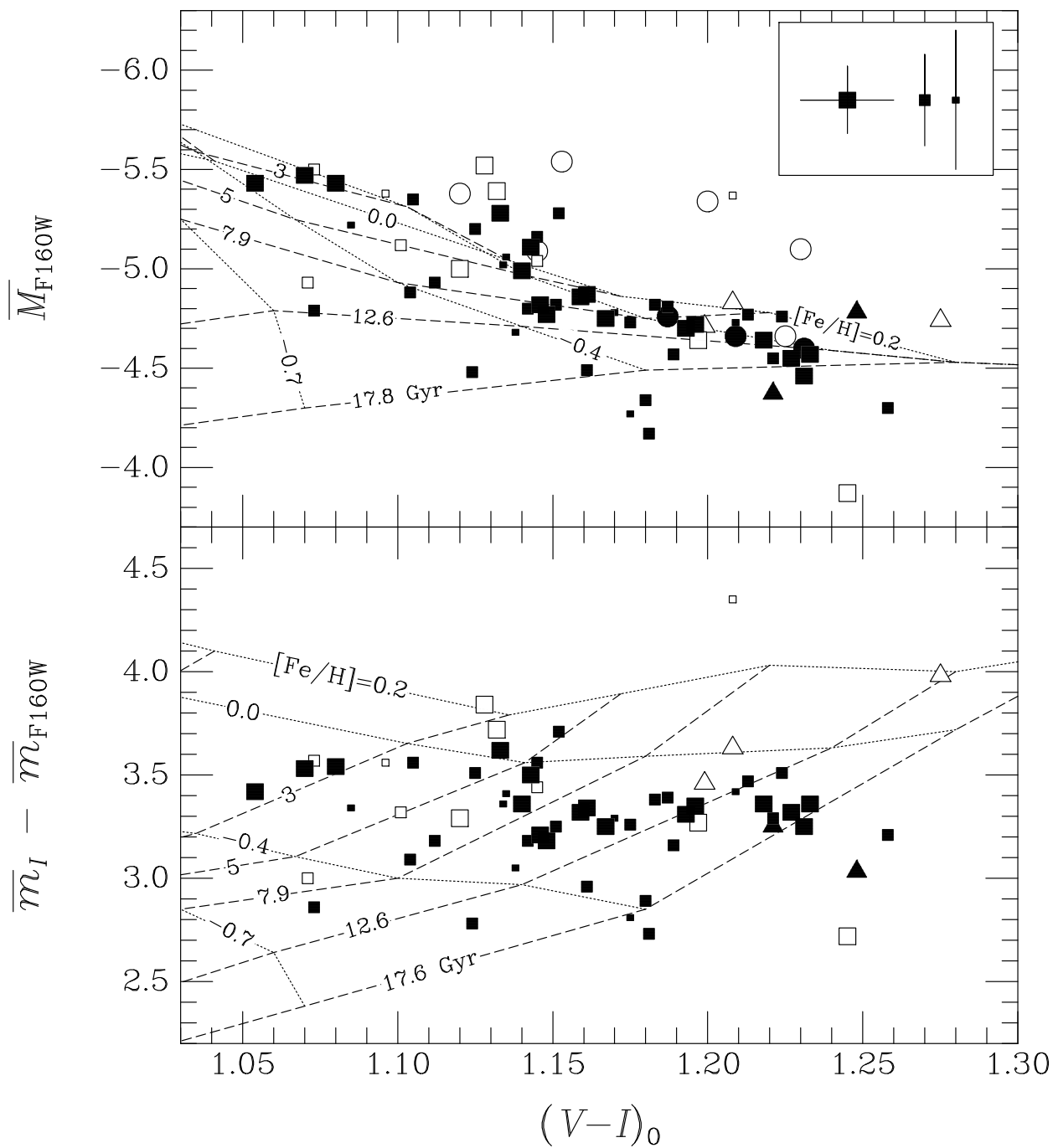}
\figcaption[]{Absolute fluctuation magnitudes and fluctuation
colors plotted with the 
Vazdekis stellar population models (Blakeslee et al. 2001), 
translated to the F160W filter as described in the text.
The symbols are the same as in Figure~\ref{liufig}.
\label{bvafig}}
\end{figure}

In Figure~\ref{bvafig} we compare our data to the F160W 
SBF magnitudes and \vminio\ colors computed for the Vazdekis 
models (Blakeslee et al. 2001; Vazdekis 2001).  
Additional models for populations younger than 4 Gyr were retrieved 
from {\tt http://www.iac.es/galeria
/vazdekis/col\_lick.html}.
The Blakeslee et al. models make use of the new Padova isochrones
of Girardi et al. (2000), which are transformed to the observable
plane using the empirical stellar libraries of Alonso, Arribas, \&
Martinez-Roger (1996, 1999)
and Lejeune et al. (1997, 1998).
The Blakeslee et al. (2001) SBF models were computed for the 
$H$-filter, so it is necessary to shift the models to F160W.
We adopted an empirical correction to the absolute fluctuation magnitudes of
\begin{equation}
\overline M_{\rm F160W} = \overline M_H + 0.10(\overline M_J{-}\overline M_K)
\label{stephenseq}
\end{equation}
based on photometry of red main sequence stars 
(Stephens et al. 2000).
This correction, which is typically 0.17~mag for most models, has the
effect of shifting the model lines in Figure~\ref{bvafig} down with
respect to the points.  
We also computed the offset between $H$ and F160W by convolving the
filter profiles with synthetic spectra of red giant stars (B. Plez,
private communication; see also Bessell, Castelli, \& Plez 1998).
The resulting relation
\begin{equation}
\overline M_{\rm F160W} = \overline M_H + 0.116(\overline M_J{-}\overline M_K) + 0.026
\end{equation}
is very close to that derived from the Stephens
et al. (2000) data and has no effect on the interpretation, 
so we adopted the empirical correction in Equation
\ref{stephenseq}.

The top panel in Figure~\ref{bvafig} reveals very good agreement
between the models as corrected using Equation~\ref{stephenseq} and
the data calibrated using the distances based on the Freedman
et al. (2001) Cepheid calibration without metallicity corrections.
Uncertainties in the distance calibration are avoided when we compare
the fluctuation color \colorbar\ to the models.  
The ages and metallicities implied by the \colorbar\ models are in very 
good agreement with those of the \Mbar\ models and with the 
Liu et al. (2002) models.
Once again, a change in the distance scale zero point of more than
approximately 0.1~mag would compromise the compatibility between
the models.  

The Vazdekis models show a compression in \Mbar\ of the 
higher-metallicity and younger-age tracks compared to the 
Liu et al. (2000, 2002) models.
This is partly due to the fact that the maximum metallicity shown 
for the Blakeslee et al. (2001) models is 0.2, while that of Liu et al. 
is 0.4.  It is also partly due to the difference in slope between
the youngest populations.  While there are many differences
in the details of the Vazdekis and Bruzual \& Charlot models, the choice 
of isochrones is perhaps the most significant for the current study.  
If we apply the newer evolutionary tracks (Girardi et al. 2000) 
adopted by Blakeslee et al. (2001) to the Bruzual \& Charlot 
models (Liu et al. 2000, 2002), we find that the
resulting age tracks are nearly horizontal and lie midway
between the two sets of models shown in Figures~\ref{liufig} and 
\ref{bvafig}.


\subsection{The Worthey Models}

\begin{figure}
\plotone{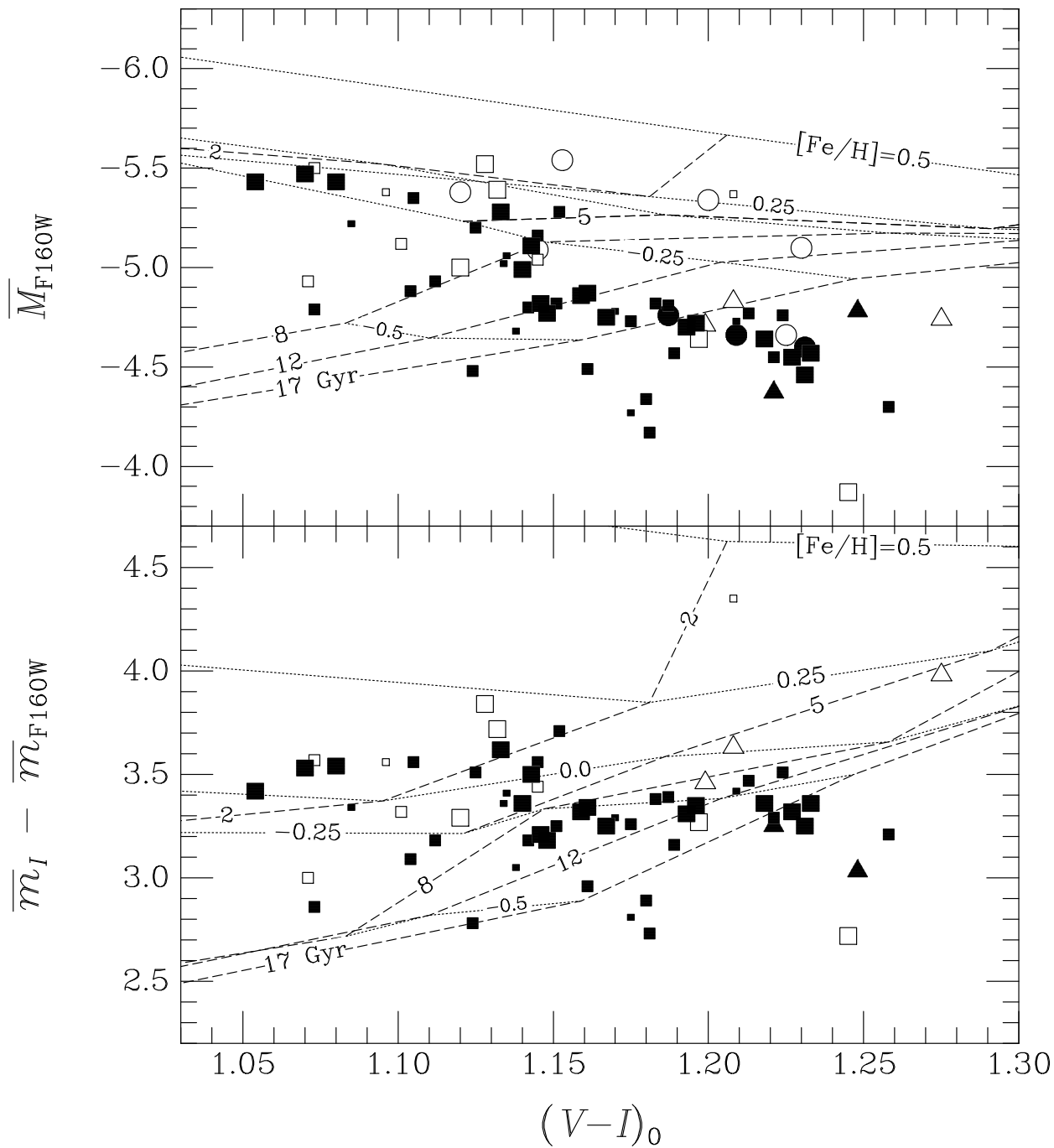}
\figcaption[]{Absolute fluctuation magnitudes and fluctuation colors
plotted with the Worthey (1994) models.  
The symbol definitions and uncertainties are the
same as in Figure~\ref{liufig}.
Note that the triangular gap on the left is caused by the
lack of published models for young, metal-poor populations, and is
not a feature of the models.
\label{wortheyfig}}
\end{figure}

Figure~\ref{wortheyfig} shows the F160W data plotted with the
Worthey (1994) models.
Worthey models are currently
available via the web at
{\tt http://astro.wsu.edu/worthey/dial/dial\_a\_model.html}.
F160W models
were constructed by convolving the basic ``vanilla'' model 
spectral energy distributions at each age and metallicity with the 
filter profile of the NIC2 F160W filter (G. Worthey, private communication).
The default Worthey models use a Salpeter initial mass function 
and a helium fraction of ${\rm Y}\,{=}\,0.228\,{+}\,2.7{\rm Z}$.
There are differences in detail between the Worthey
models and those presented in the previous sections.  
In particular, the Worthey models
imply older ages for galaxies of a given color, and unreasonably
large ages for the reddest galaxies.  
Note that the calibration changes to the Freedman et al. (2001)
Cepheid zero point without metallicity corrections makes the
comparisons to the Worthey \Mbar\ models worse than they would have
been using the previous Key Project calibration (Ferrarese et al. 2002b,
J2001).
The comparison of the distance-independent \colorbar\ models 
to the data yields ages and metallicities in excellent agreement with 
the conclusions of the previously-considered models.  
The redder galaxies have slightly sub-solar metallicities and old
ages, while the bluer galaxies appear to have higher metallicities
and significantly younger ages.  
It appears that the Worthey (1994) models may have an offset in
absolute magnitude, but the relative brightnesses
of the models at different wavelengths are reliable.


\subsection{Composite Stellar Population Models}

Fluctuation magnitudes are dominated by the youngest stars in a population.
Statistically, measured SBFs are the ratio of the second moment 
of the stellar luminosity function to the first, 
and therefore are very sensitive to the brightest stars.  When comparison
of SBF magnitudes to the models suggests the presence of young
or intermediate-age populations (${<}5$ Gyr), only a fraction of the 
stars would have to be young for the fluctuation amplitude to be high; 
the majority of the stars may be much older.

Tonry, Ajhar, \& Luppino (1990) were the first to attempt to use 
composite stellar
population models to explain optical SBF observations, but the inadequacy of
their model isochrones and colors led Worthey (1993) to question the
validity of their results.  More recently, Blakeslee et al. (2001)
and Liu et al. (2002) have compared composite stellar population models
to observational SBF data.  Blakeslee et al. found that a 3-component
model that includes a ${\sim}10\%$ metal-poor component reproduces the 
observed $I$-band SBF slope and agrees with the IR SBF results presented
here.

\begin{figure}
\plotone{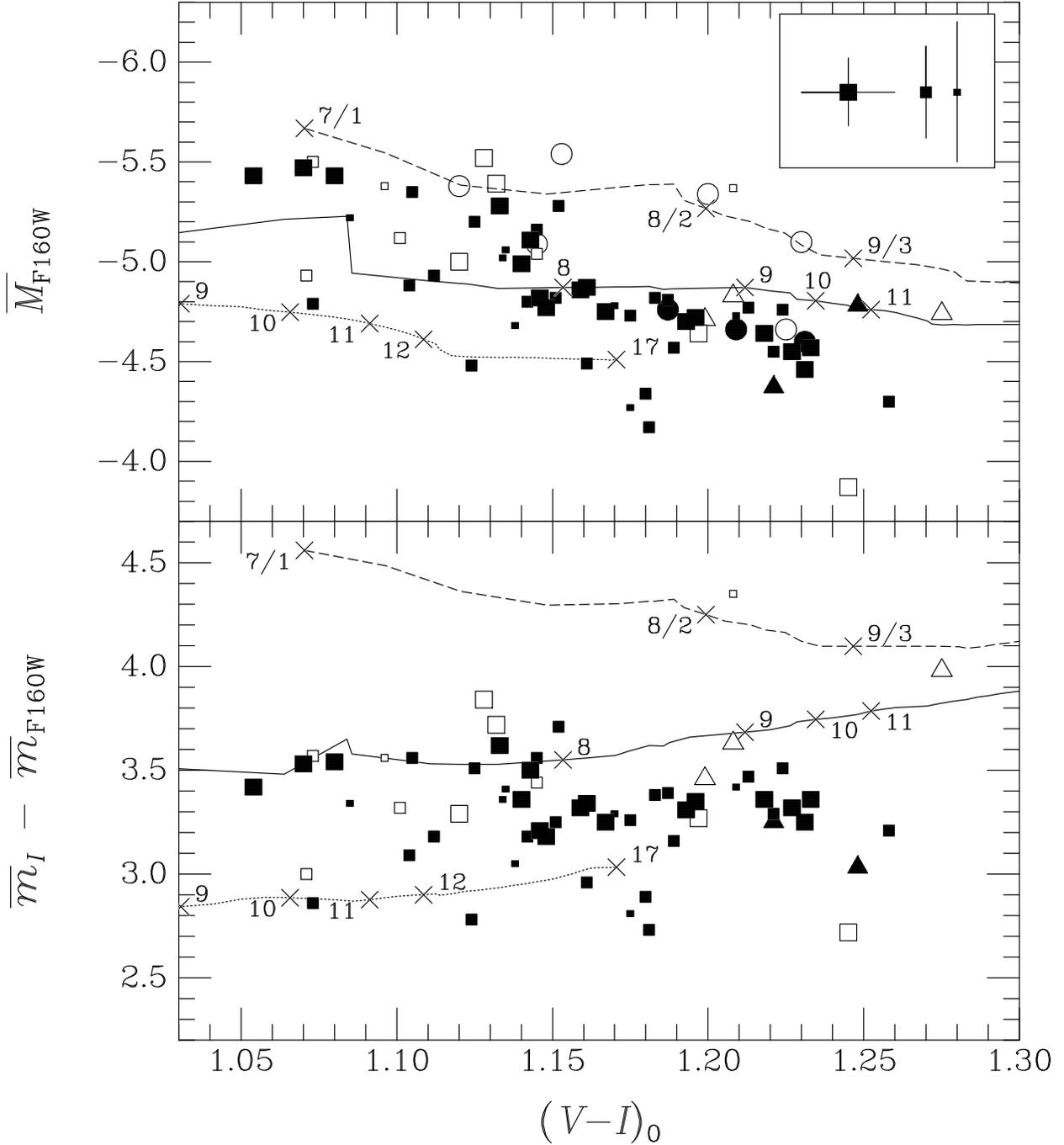}
\figcaption[]{
Three composite stellar population models (Liu et al. 2002) 
compared to the SBF measurements.  In each case, the model population
is composed of two bursts separated by 6 Gyr, with 
the younger population containing 20\% of the total mass.
The total age of the composite population in Gyr is indicated by the 
numbers:  8 indicates the position on the line of a population
where 80\% of the stars by mass formed 8 Gyr ago and 20\% formed
2 Gyr ago.  
The top (dashed) line indicates models with the second burst forming
from super-solar metallicity gas ([Fe/H]$\,{=}\,0.4$), the center (solid) 
line from solar metallicity gas ([Fe/H]$\,{=}\,0.0$), and the lowest line 
(dotted) from sub-solar metallicity gas ([Fe/H]$\,{=}\,{-}0.7$).  
In all cases the older population is of solar metallicity.  
\label{compositefig}}
\end{figure}

In Figure~\ref{compositefig} we compare our fluctuation magnitudes to 
the composite stellar population models of Liu et al. (2002)
re-calculated for the F160W filter.  In the three scenarios shown
in Figure~\ref{compositefig}, 20\% of the final mass formed 
six Gyr after the principal population.  Composite populations with
smaller mass fractions and smaller age differences would fall between 
the tracks shown in Figure~\ref{compositefig}; these models are 
meant to be representative and are shown 
for comparison with those published by Liu et al. (2002).
The total time since formation is indicated in Figure~\ref{compositefig}
on each evolutionary track,
thus the 7~Gyr point marks the magnitude and color of a population
where 80\% of the stars by mass are 7~Gyr old and the other 20\% formed 
1 Gyr ago.  The top line is for a burst
with higher metallicity ([Fe/H]$\,{=}\,{+}0.4$) than the original 
solar-metallicity population.  The middle track is for a burst
population of solar metallicity, and the lowest track is for a burst
with lower-than-solar metallicity ([Fe/H]$\,{=}\,{-}0.7$).
In all three cases the majority of the stars (80\% of the final mass) 
formed from solar-metallicity gas.

Figure~\ref{compositefig} shows that the bright fluctuations in
bluer galaxies may be the result of a recent burst of star formation
within the last $\sim$2 Gyr with a total mass fraction of ${\sim}20$\%.
Liu et al. (2002) came to the same conclusion comparing $K$-band
SBF data to the composite models.
The F160W data in Figure~\ref{compositefig} suggest 
that the burst took place in previously-enriched gas.
If recent episodes of star formation are triggered by mergers with 
dwarf galaxies with lower than solar metallicity, the star formation
must take place in gas previously enriched in the larger galaxy.
Figure~\ref{compositefig} suggests that as time goes on, bursts of
star formation take place in increasingly metal-rich gas.

The bottom panel of Figure~\ref{compositefig} shows the comparison
of fluctuation color \colorbar\ to the composite models.  
The \colorbar\ models imply slightly lower absolute metallicities 
for galaxies at the blue end of the color distribution than the \Mbar\ 
models do.  
None of the points in the lower panel fall much above the solar metallicity
\colorbar\ track, in contrast to the data in the top panel.
The metallicities implied for the redder galaxies are the same
in both panels.
The metallicities implied by the \colorbar\ models support the conclusion
that more-recent star formation takes place in progressively 
higher-metallicity gas. 


\subsection{Comparing Fluctuations to the \hbeta\ Index\label{hbetasec}}

The well-known age-metallicity degeneracy makes it difficult to
use broad-band optical colors to distinguish old, metal-poor
populations from younger, metal-rich ones (e.g., Worthey 1994).
One currently popular technique for determining ages and metallicities, 
particularly in young populations ($t{<}5$ Gyr), is to compare 
the age-sensitive \hbeta\ absorption line strength to a metal line 
that is sensitive to metallicity.  
IR SBFs also break the age-metallicity degeneracy and allow
one to distinguish young, blue populations from older, metal-poor
populations (Liu et al. 2000; Blakeslee et al. 2001).  
SBFs are dominated by the brightest stars in a galaxy,
which in relatively young and intermediate-age populations 
($1\,{<}\,t\,{<}\,5$ Gyr) are evolved red giant and AGB stars.  
The populations we are considering are not 
actively forming stars, and are not so young as to contain massive 
luminous OB stars or emission nebulae.  
In contrast to SBFs, the \hbeta\ absorption arises from main sequence 
stars near the main sequence turn-off.  Because \hbeta\
and SBF measurements arise from {\it completely different} groups of 
stars, and are measured using {\it completely different} techniques, the
F160W SBF measurements presented here provide an important independent
confirmation that \hbeta\ absorption is truly revealing differences
in the ages of stellar populations.  F160W fluctuation
magnitudes and \hbeta\ absorption are both sensitive probes of
young stellar populations. 
The fact that the two techniques agree for individual galaxies in
this sample is significant.

\begin{figure}
\plotone{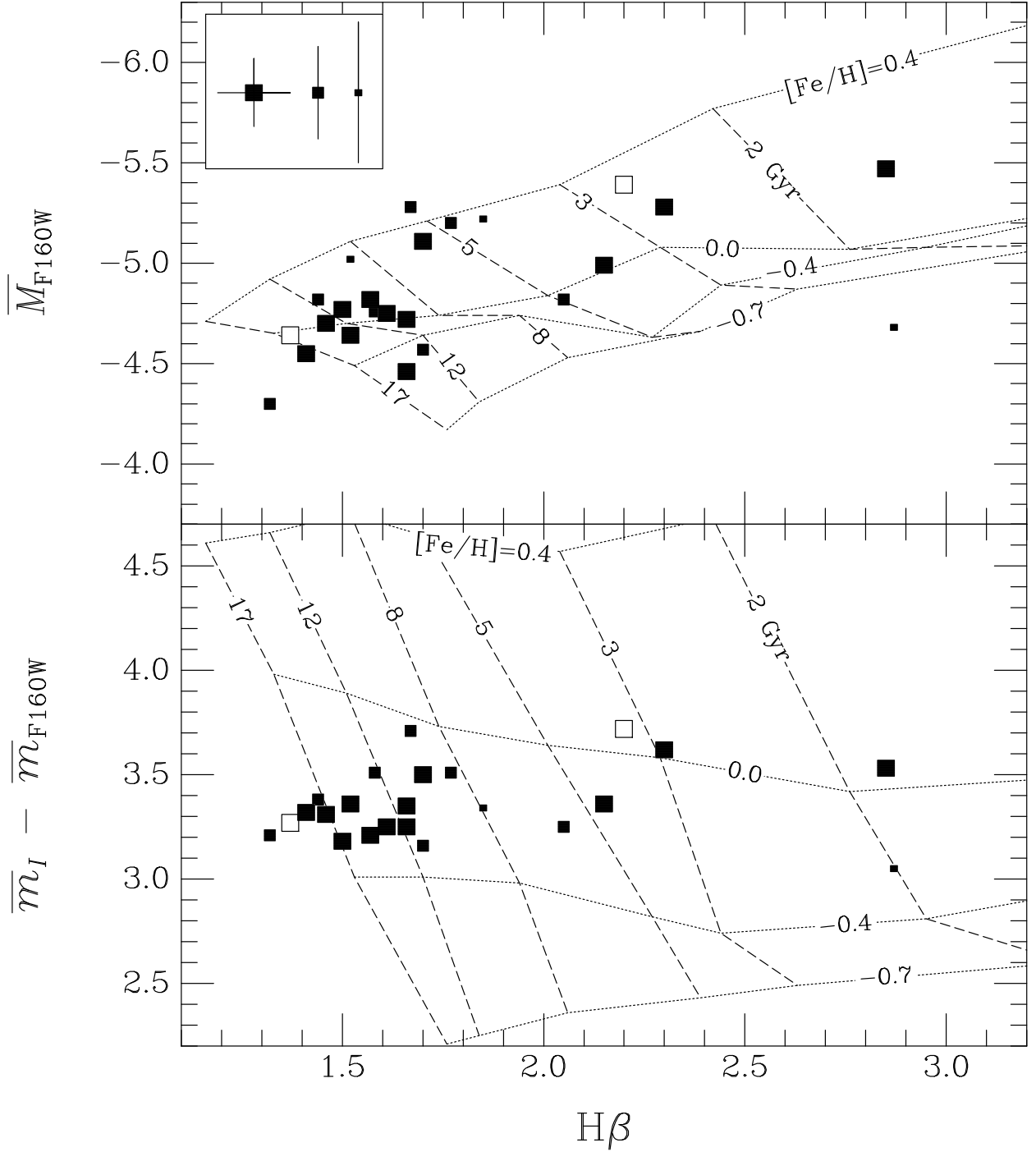}
\figcaption[]{
Absolute fluctuation magnitudes and fluctuation colors 
plotted vs. the \hbeta\ absorption line index (in \AA) 
for a subset of 24 galaxies (Kuntschner et al. 2000,
and Kuntschner 2000, and Trager et al. 2000).
The majority of the galaxies are ellipticals in the Fornax cluster
(see Table~\ref{observationtable}).
\hbeta, like \Mbar, is sensitive to the age of the stellar population.
Model predictions of both \Mbar\ and \hbeta\ are shown for the
Liu et al. (2000, 2002) models, with ages and metallicities indicated.
The symbol definitions are the same as in Figure~\ref{liufig}.
\label{hbetafig}} 
\end{figure}

In Figure~\ref{hbetafig} we compare the age-sensitive
\Mbar\ to \hbeta\ and use the Bruzual \& Charlot models of 
Liu et al. (2002) to infer relative ages and metallicities.  
We collected high-quality \hbeta\ measurements 
on the Lick/IDS system from three recent studies (Kuntschner et al. 
2000; Kuntschner 2000; Trager et al. 2000).  These measurements
include a sample of 14 galaxies in the Fornax cluster and 10 others, 
including some in the Local Group, Virgo, Leo clusters. 
The three studies measured \hbeta\ values in apertures of different sizes.
Kuntschner (2000) used a single aperture size for the Fornax cluster
galaxies.  Kuntschner et al. (2000) scaled their measurements by
distance to a common physical scale.  The Trager et al. (2000) data 
are measured in an aperture scaled to the size of the galaxy ($r_e/8$).
It is unnecessary to adjust the measurements to a common system since
Kuntschner et al. (2000) found that \hbeta\ does not change significantly
with radius from the galaxy center.  
The Kuntschner and Trager studies apply a slightly 
different correction for \hbeta\ emission; we applied a small 
correction of 0.1 times [OIII]${\lambda}5007$ to the Trager et al. data 
to match the correction used in the Kuntschner papers.  
The \hbeta\ measurements are listed in Table~1.
Two galaxies were measured by both teams:  \hbeta\ measurements for
NGC~3379 are presented in all three papers, and data for NGC~4472
are published in two.  The \hbeta\ values listed in Table~1 are 
averages of all measurements, and the uncertainties were added
in quadrature.  The standard deviation of the different measurements
was less than or equal to the total uncertainties in both cases.

The trend seen in Figure~\ref{hbetafig} shows that the galaxies span 
a wide range in age, both as determined using \hbeta\ and \Mbar.
There is clearly a correlation between \Mbar\ and \hbeta.
The stellar population models (Liu et al. 2002) have lines of constant
age perpendicular to the correlation between \Mbar\ and \hbeta\ seen
in Figure~\ref{hbetafig}, indicating that both are sensitive to 
young and intermediate-age populations 
(even though they sense very different {\em components} 
of the young populations).
The fact that both techniques lead to the same conclusion that early-type
galaxies often contain young stellar populations strengthens the
conclusion considerably.
The stellar population models become somewhat more compressed for 
ages greater than 8 Gyr, when the A stars responsible for 
\hbeta\ absorption have evolved off the main sequence, and the red
giant and AGB stars responsible for the fluctuations are intrinsically 
fainter and more uniform in brightness (the model ages are best interpreted 
in a relative sense; observations that agree with the oldest models do not 
imply ages older than the Universe).

The absolute metallicities
implied by Figure~\ref{hbetafig} are only marginally higher than those 
seen in the previous comparisons, and the trend with age is consistent.
In the lower panel of Figure~\ref{hbetafig} the \hbeta\ values are
compared to the fluctuation color \colorbar,  which is distance
independent.  
The \Mbar\ and \colorbar\ models disagree slightly.  
The fluctuation color models suggest 
somewhat lower metallicities than the \Mbar\ models,
in agreement with the other model comparisons.

Both panels in Figure~\ref{hbetafig} show a relationship between
IR fluctuations and \hbeta.  The models suggest that both elliptical
and S0 galaxies span a very wide range in age and a relatively narrow 
range of metallicity.
This result disagrees somewhat with the conclusion of Kuntschner (2000),
who found that only the S0 galaxies in Fornax have a wide spread in
age.
Kuntschner's limited sample of Fornax ellipticals appear uniformly old 
and span a range in metallicity.
Our results are consistent with the conclusions of Trager 
et al. (2000), who found a wide age spread among a sample of 
ellipticals drawn from a variety of environments.


\subsection{Common Ground}

In detail, there are significant differences between the different 
models in predicted fluctuation magnitudes.  
Differences have also been noted at other wavelengths
(Blakeslee et al. 2001, Liu et al. 2002, Mei et al. 2001b).
In general, however, the comparison of empirical fluctuation
magnitudes to the models yields a broadly consistent picture.  
Fluctuation amplitudes in galaxies redder than \vminio$\,{>}\,1.16$ 
are all consistent with stellar populations older than approximately 
8~Gyr and of slightly sub-solar metallicity.
Galaxies on the bluer end of the color range all appear consistent with the 
youngest (2 to 5 Gyr) models in both absolute fluctuation magnitude and
fluctuation color.  
Regardless of which model is considered, these blue, lower-luminosity 
galaxies appear to be more metal-rich than the redder ellipticals.
None of the bluer
galaxies in this sample have fluctuation magnitudes that are 
consistent with old (${>}5$ Gyr) 
single-burst stellar population models with low metallicities.  
The majority of stars in the blue
galaxies may still be old, as the fluctuation magnitudes are dominated 
by young stars in a population.
Composite stellar population models reinforce the conclusions of the
single-age stellar population models.

The conclusion that bluer early-type galaxies are younger and more 
metal-rich seems to contradict the observed relationship between mass and 
metallicity in giant ellipticals where less massive (bluer) systems retain
less enriched gas, and should therefore have lower metallicities
than the massive ellipticals.
The SBF magnitudes are dominated by the brightest and youngest stars
in a galaxy, however, so our conclusion regarding metallicity
may be consistent with the mass--metallicity relation if the 
bluer ellipticals have lower metallicities overall and some
small fraction of their stars formed relatively recently from
more metal-rich gas.  
The uncertainty in the assumed luminosity evolution of the AGB stars in the 
youngest population models also serves to minimize the importance of
the apparent contradiction.  
A modest excess in the real AGB population over the models would 
result in enhanced SBF magnitudes that would appear the same as 
enhanced metallicities. 

The differences between models indicate that the details of the 
evolutionary tracks that are used in the models deserve further 
attention.
Furthermore, the Padova evolutionary tracks are not computed
beyond the onset of the thermally-pulsating stage of the AGB, and the
modelers have employed a synthetic parameterization of the
evolution of AGB stars to attempt to
reproduce the luminosity function of realistic populations.
The data presented here can provide feedback that should
result in improved understanding of the luminosities and lifetimes of
luminous AGB stars.  It is also clear that real galaxies are not 
composed of coeval populations of stars of constant metallicity.  

The excellent agreement between the data and the models for both 
the \Mbar\ and the distance-independent \colorbar\ models suggests 
that the Cepheid distances are probably reliable to better than 10\%.
For the current study we adopted the new Cepheid
period-luminosity relations of Udalski et al. (1999), no 
metallicity corrections, and a distance modulus to the LMC of 18.50~mag.

\section{Galaxy Morphologies and SBF Magnitudes}

\begin{figure}
\plotone{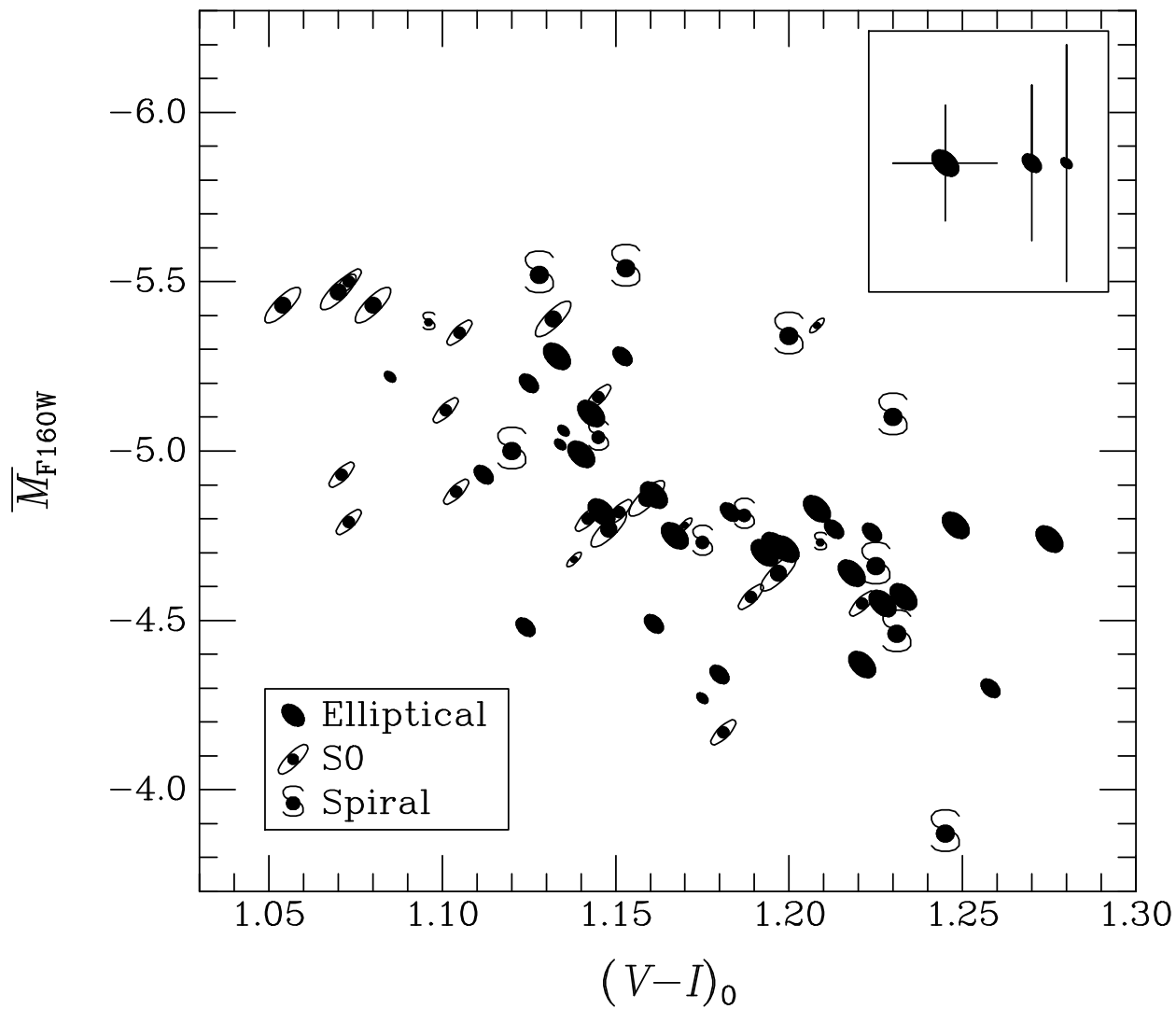}
\figcaption[]{
Absolute fluctuation magnitudes plotted vs. color by 
galaxy type.  T-types (RC3) of $-4$ to $-6$ are plotted as ellipticals, 
$-3$, $-2$, and $-1$ as lenticular (S0) galaxies, and types
T${\ge}0$ are shown as spirals.  As in the previous figures, 
the largest symbols have the smallest uncertainties (median 
uncertainties are indicated in the upper right corner).
\label{ttypefig}}
\end{figure}

\begin{figure}
\plotone{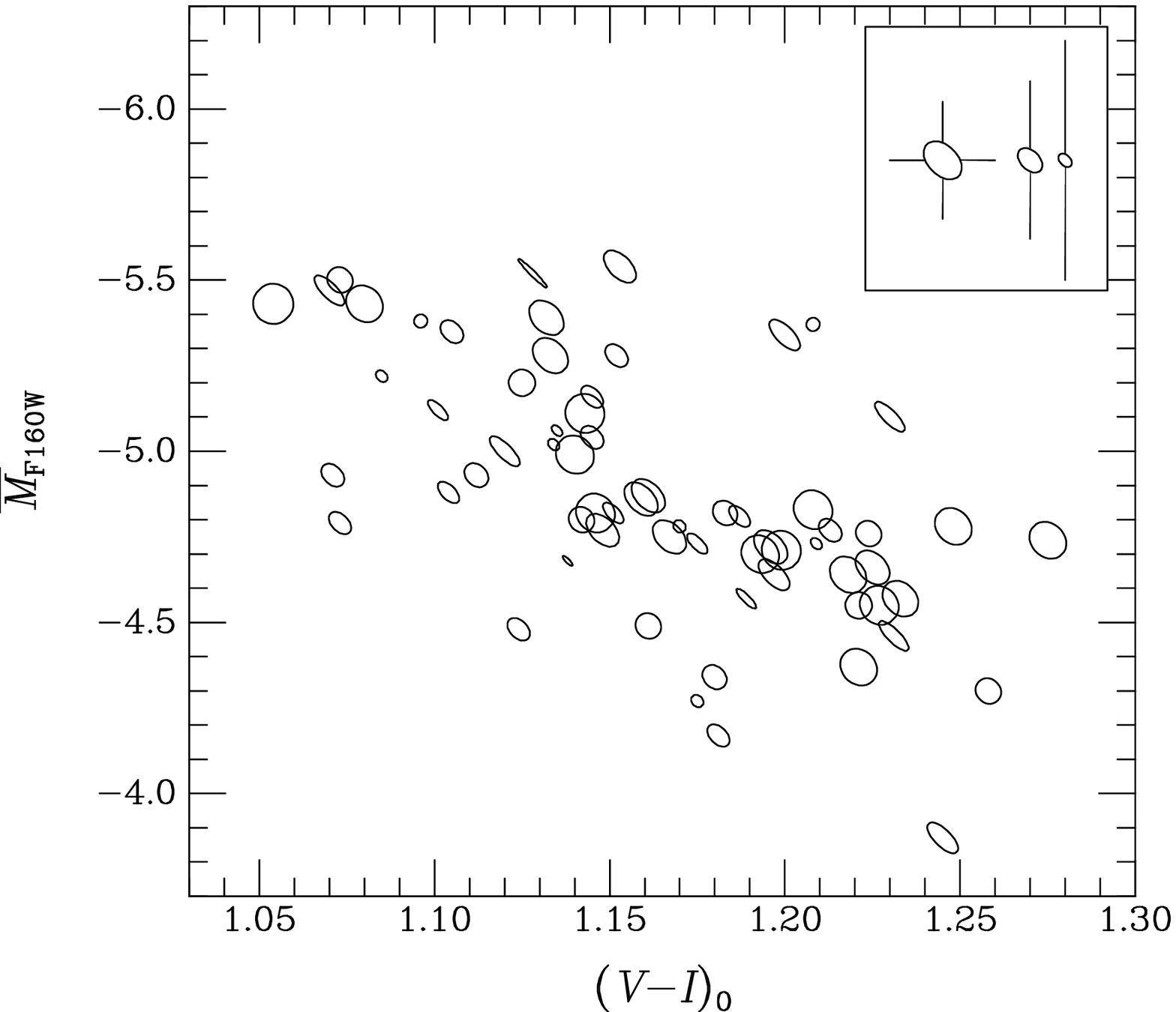}
\figcaption[]{
Absolute fluctuation magnitudes plotted vs. color, with 
symbol shapes that indicate the projected ellipticity on the sky
(RC3).  
\label{ellipfig}}
\end{figure}

Figure~\ref{ttypefig} shows the absolute fluctuation magnitudes 
separated into bins by morphological type.
Galaxy morphology classifications were taken from de Vaucouleurs
et al. (1991, RC3).  Absolute fluctuation
magnitudes and the projected ellipticities of the galaxies on the sky
are plotted in Figure~\ref{ellipfig} (RC3).  The shapes of the
symbols in Figure~\ref{ellipfig} are the same as the optical
shapes of the galaxies measured well outside the NICMOS field of
view.  Morphological types and ellipticities from RC3 are
listed in Table~\ref{observationtable}.

As a group, lenticular S0 galaxies are
bluer than the elliptical population; nevertheless, the overlap
between elliptical and S0 galaxies covers almost the entire
range in color covered by this sample.  The bulges of spiral
galaxies also span the full range of color and \Mbar\
spanned by the elliptical galaxies.  Some dusty spiral bulges have
fluctuation magnitudes that are significantly brighter than
the early-type galaxies of the same color, most likely because
the fluctuation amplitude is enhanced by the presence of clumpy
dust.
Some galaxies of all types included in our sample 
(T$\,{=}\,{-}6$ to $+4$) apparently contain bluer and brighter
intermediate-age populations near their centers.
While most of the galaxies are found in clusters or groups, 
the sample covers a range of environment from dense, compact 
clusters like Fornax, to lower-density groups and clusters (Leo, 
Virgo, and Ursa Major).  Thirteen galaxies are not associated 
with any obvious group or cluster (as defined by Faber et al. 1989). 


\section{Summary}

1.  Using F160W SBF magnitudes to accurately determine extragalactic
distances requires that the color of the galaxy be known.  
The best calibration fit is given in Equation~\ref{calibeq}.
To achieve an accuracy of ${\sim}10$\% or better, the uncertainty in 
\vminio\ color should be less than ${\sim}0.035$~mag.
Applying the new calibration to the distant F160W SBF data of J2001
would result in a value of the Hubble constant that is 10\%
larger than that published previously; almost all of the
difference (8\%) is due to the improved Cepheid period-luminosity 
relation of Udalski et al. (1999). 

2.  The reddest, most-massive ellipticals appear older (when
compared to theoretical stellar population models) than the 
less-luminous bluer galaxies.  
There are {\it no} galaxies in the blue half of our sample that have 
fainter fluctuation magnitudes consistent with very low metallicity
([Fe/H]$\,{<}\,{-}0.7$) and old age ($t{>}5$ Gyr) stellar population models.
Comparison with stellar population models suggests that the 
youngest galaxies have somewhat higher metallicities than
the older ellipticals.  None of the galaxies have fluctuation
magnitudes that are too bright to be accommodated by the young,
metal-rich models. 

3.  Composite stellar population models composed of ${\sim}20$\% 
younger stars by mass are also consistent with the brighter fluctuation 
magnitudes in bluer galaxies, provided the younger stars formed from gas 
of equal or higher metallicity than the gas that formed the older
stellar population.

4.  The \hbeta\ line index and fluctuation magnitude \Mbar\ are both 
sensitive to young and intermediate-age stellar populations with ages 
between one and a few Gyr.  
Age estimates from the two techniques are consistent.
The agreement is significant because the two types of measurements 
make use of {\it completely different} techniques and are sensitive to 
{\it different populations} of stars.

5.  Comparison of fluctuation magnitudes and distance-independent 
fluctuation colors to the predictions of stellar population
models provides an independent check on the distance scale calibration.  
If the models are correct, at least in the relative brightnesses 
between the $I$ and F160W bands, then the total systematic error
in the Cepheid distance scale is possibly ${\lesssim}0.1$ mag.

6.  The S0 galaxies and dust-free spiral bulges in this sample 
have fluctuation magnitudes that are indistinguishable from those of the
elliptical galaxies of the same color.  Young and intermediate-age
populations of relatively high metallicity must therefore be present
in all galaxy types on the blue end of the color distribution.
While there is a greater fraction of S0 galaxies among the bluer
galaxies, galaxies of all types are found across the entire color
range.

\acknowledgements
This study benefited greatly from NICMOS data collected as part of several 
programs, and we thank those who worked to acquire that data.  
In particular, we are grateful to those who helped ensure that the
data would be appropriate for SBF analysis and assisted with the
data reductions (D. Geisler, J. Elias, J. R. Graham, and S. Charlot).
We greatly appreciated the helpful comments of 
L. Ferrarese, T. Lauer, M. Postman, and R. Weymann.
This study made use of data collected by the Optical SBF
team (J. Tonry, J. Blakeslee, E. Ajhar, and A. Dressler), and we thank
them for providing color photometry and $I$-band SBF data.  
Finally, we wish to thank S. Charlot, A. Vazdekis, and G. Worthey for 
constructing the SBF stellar population models.  

This research was supported in part by NASA grant GO-07453.0196A.  
The NICMOS GTO team was supported by NASA grant NAG 5-3042.
J. Jensen acknowledges the support of the Gemini Observatory,
which is operated by the Association of Universities for Research in
Astronomy, Inc., on behalf of the international Gemini partnership of 
Argentina, Australia, Brazil, Canada, Chile, the United Kingdom, and the 
United States of America.  M. Liu acknowledges the support of a
Beatrice Watson Parrent Fellowship.

\clearpage

\begin{deluxetable}{llclrrrrrrrc}
\tabletypesize{\scriptsize\tiny}
\tablecolumns{11}
\tablecaption{F160W NIC2 Observational Data\label{observationtable}}
\tablehead{
\colhead{Galaxy} &
\colhead{Grp.\tablenotemark{a}} &
\colhead{Prog.} &
\colhead{Obs.} &
\multicolumn{2}{c}{Position (J2000)} &
\colhead{$t_{\rm exp}$} &
\colhead{$A_{\rm B}$ \tablenotemark{b}} &
\colhead{c$z$} &
\colhead{T \tablenotemark{c}} &
\colhead{$e$ \tablenotemark{c}} &
\colhead{\hbeta\ \tablenotemark{d}} \\
& & 
\colhead{ID} &
\colhead{dataset} &
\colhead{RA}&
\colhead{dec} &
\colhead{(s)} &
\colhead{(mag)} &
\colhead{(\kms)} & & &
\colhead{(\AA)} 
}
\startdata
NGC 7814 &\nodata& 7330 & N3ZB1A & 00:03:15.12 &    16:08:49.7 &  640 & 0.194 & 1054 &   +2 & 0.58 &\nodata   \\
NGC 221  & 282   & 7171 & N4EY01 & 00:42:41.85 &    40:51:51.8 &   64 & 0.268 &$-$145& $-$6 & 0.26 &$2.30{\pm}0.05$   \\
NGC 224  & 282   & 7171 & N4EYA1 & 00:42:44.33 &    41:16:08.4 &   64 & 0.268 &$-$300&   +3 & 0.68 &$1.66{\pm}0.07$   \\
NGC 404  & 282   & 7330 & N3ZB2L & 01:09:26.80 &    35:43:05.3 &  640 & 0.253 &$-$48 & $-$3 & 0.00 &\nodata   \\
NGC 524  &\nodata& 7886 & N4RW05 & 01:24:47.74 &    09:32:19.8 &  640 & 0.357 & 2421 & $-$1 & 0.00 &\nodata   \\
NGC 708  & 27    & 7453 & N4HD14 & 01:52:46.49 &    36:09:06.5 &  960 & 0.379 & 4871 & $-$5 & 0.17 &\nodata   \\
NGC 821  &\nodata& 7886 & N4RW27 & 02:08:21.15 &    10:59:42.0 &  640 & 0.478 & 1718 & $-$5 & 0.37 &$1.66{\pm}0.04$   \\
NGC 1052 & 207   & 7886 & N4RW09 & 02:41:04.75 & $-$08:15:20.7 &  640 & 0.115 & 1507 & $-$5 & 0.31 &\nodata   \\
NGC 1172 & 29    & 7886 & N4RW10 & 03:01:36.04 & $-$14:50:12.0 &  640 & 0.290 & 1550 & $-$4 & 0.22 &\nodata   \\
NGC 1316 & 31    & 7458 & N4B707 & 03:22:41.51 & $-$37:12:33.0 &  768 & 0.090 & 1760 & $-$2 & 0.29 &$2.20{\pm}0.07$   \\
NGC 1351 & 31    & 7886 & N4RW15 & 03:30:35.01 & $-$35:51:14.2 &  640 & 0.061 & 1518 & $-$3 & 0.40 &$1.50{\pm}0.10$   \\
NGC 1339 & 31    & 7458 & N4B710 & 03:28:06.58 & $-$32:17:04.3 &  256 & 0.057 & 1392 & $-$4 & 0.28 &$1.52{\pm}0.11$   \\
NGC 1344 & 31    & 7458 & N4B709 & 03:28:18.98 & $-$31:04:04.3 &  256 & 0.077 & 1169 & $-$5 & 0.42 &\nodata   \\
NGC 1365 & (31)  & 7330 & N3ZB30 & 03:33:35.36 & $-$36:08:22.2 &  640 & 0.088 & 1662 &   +3 & 0.45 &\nodata   \\
NGC 1373 & 31    & 7458 & N4B7A3 & 03:34:59.25 & $-$35:10:17.0 &  256 & 0.060 & 1334 & $-$4 & 0.26 &$1.85{\pm}0.10$   \\
NGC 1374 & 31    & 7458 & N4B703 & 03:35:16.66 & $-$35:13:34.3 &  256 & 0.060 & 1294 & $-$5 & 0.07 &$1.57{\pm}0.09$   \\
NGC 1375 & 31    & 7458 & N4B706 & 03:35:16.84 & $-$35:15:56.5 &  256 & 0.063 &  740 & $-$2 & 0.62 &$2.85{\pm}0.09$   \\
NGC 1379 & 31    & 7453 & N4HD07 & 03:36:04.03 & $-$35:26:26.8 &  384 & 0.052 & 1324 & $-$5 & 0.05 &$1.70{\pm}0.09$   \\
NGC 1380 & 31    & 7458 & N4B702 & 03:36:27.15 & $-$34:58:33.4 &  265 & 0.075 & 1877 & $-$2 & 0.52 &$1.37{\pm}0.11$   \\
NGC 1381 & 31    & 7458 & N4B7A6 & 03:36:31.90 & $-$35:17:46.4 &  256 & 0.058 & 1724 & $-$2 & 0.72 &$1.70{\pm}0.06$   \\
NGC 1386 & 31    & 7458 & N4B708 & 03:36:45.37 & $-$35:59:57.0 &  256 & 0.054 &  868 & $-$1 & 0.62 &\nodata   \\
NGC 1380A& 31    & 7453 & N4HDA7 & 03:36:47.62 & $-$34:44:25.2 &  384 & 0.063 & 1561 & $-$2 & 0.71 &$2.87{\pm}0.13$   \\
NGC 1387 & 31    & 7458 & N4B701 & 03:36:57.06 & $-$35:30:22.7 &  256 & 0.055 & 1302 & $-$3 & 0.00 &\nodata   \\
NGC 1389 & 31    & 7458 & N5B7A8 & 03:37:11.68 & $-$35:44:45.5 &  256 & 0.046 &  986 & $-$3 & 0.40 &\nodata   \\
NGC 1399 & 31    & 7453 & N4HD09 & 03:38:29.09 & $-$35:27:00.6 &  384 & 0.058 & 1425 & $-$5 & 0.07 &$1.41{\pm}0.08$   \\
NGC 1404 & 31    & 7453 & N4HDA9 & 03:38:52.04 & $-$35:35:38.3 &  384 & 0.049 & 1947 & $-$5 & 0.11 &$1.58{\pm}0.08$   \\
NGC 1400 & 32    & 7886 & N4RW17 & 03:39:30.81 & $-$18:41:16.1 &  640 & 0.280 &  558 & $-$3 & 0.13 &\nodata   \\
NGC 1427 & 31    & 7458 & N4B704 & 03:42:19.48 & $-$35:23:33.8 &  256 & 0.048 & 1388 & $-$4 & 0.32 &$1.67{\pm}0.05$   \\
NGC 1426 & 32    & 7886 & N4RW18 & 03:42:49.09 & $-$22:06:29.2 &  640 & 0.070 & 1422 & $-$5 & 0.37 &\nodata   \\
IC 2006  & 31    & 7458 & N4B705 & 03:54:28.53 & $-$35:57:54.8 &  256 & 0.048 & 1364 & $-$5 & 0.15 &$1.44{\pm}0.10$   \\
NGC 1553 & 211   & 7886 & N4RW21 & 04:16:10.28 & $-$55:46:50.8 &  640 & 0.065 & 1080 & $-$2 & 0.37 &\nodata   \\
NGC 3032 &\nodata& 7330 & N3ZB82 & 09:52:08.03 &    29:14:08.3 &  640 & 0.072 & 1533 & $-$2 & 0.11 &\nodata   \\
NGC 3056 &\nodata& 7886 & N4RW29 & 09:54:32.79 & $-$28:17:53.0 &  640 & 0.386 & 1017 & $-$1 & 0.37 &\nodata   \\
NGC 3031 &\nodata& 7331 & N3ZD0N & 09:55:32.70 &    69:03:54.0 &  384 & 0.347 &$-$34 & $+$2 & 0.48 &\nodata   \\
NGC 3351 & (57)  & 7330 & N3ZB1I & 10:43:58.08 &    11:42:16.6 &  640 & 0.120 &  778 &   +3 & 0.32 &\nodata   \\
NGC 3368 & 57    & 7330 & N3ZB2N & 10:46:45.87 &    11:49:13.6 &  640 & 0.109 &  897 &   +2 & 0.31 &\nodata   \\
NGC 3379 & 57    & 7453 & N4HD01 & 10:47:49.56 &    12:34:53.0 &  384 & 0.105 &  920 & $-$5 & 0.11 &$1.46{\pm}0.16$   \\
NGC 3384 & 57    & 7453 & N4HDA1 & 10:48:16.88 &    12:37:44.2 &  384 & 0.115 &  735 & $-$3 & 0.54 &$2.05{\pm}0.11$   \\
NGC 3928 & 155   & 7331 & N3ZD0A & 11:51:47.70 &    48:41:01.8 &  256 & 0.084 &  982 &   +3 & 0.00 &\nodata   \\
NGC 4143 & 155   & 7330 & N3ZB95 & 12:09:36.10 &    42:32:01.2 &  640 & 0.055 &  784 & $-$2 & 0.37 &\nodata   \\
NGC 4150 & 55    & 7886 & N4RW39 & 12:10:33.69 &    30:24:06.0 &  640 & 0.078 &  244 & $-$2 & 0.31 &\nodata   \\
NGC 4261 & 150   & 7868 & N4RV02 & 12:19:23.22 &    05:49:31.0 &  192 & 0.076 & 2210 & $-$5 & 0.11 &$1.32{\pm}0.06$  \\
NGC 4278 & 55    & 7886 & N4RW42 & 12:20:06.85 &    29:16:50.7 &  640 & 0.129 &  649 & $-$5 & 0.07 &\nodata   \\
NGC 4291 & 98    & 7886 & N4RW69 & 12:20:17.80 &    75:22:14.3 &  640 & 0.160 & 1757 & $-$5 & 0.17 &\nodata   \\
NGC 4406 & 56    & 7453 & N4HD03 & 12:26:11.75 &    12:56:47.7 &  384 & 0.128 &$-$227& $-$5 & 0.35 &$1.61{\pm}0.16$   \\
NGC 4434 & 56    & 7453 & N4HDA5 & 12:27:36.68 &    08:09:15.9 &  384 & 0.096 & 1071 & $-$5 & 0.02 &$1.77{\pm}0.19$   \\
NGC 4458 & 56    & 7453 & N4HDA3 & 12:28:57.59 &    13:14:31.3 &  384 & 0.103 &  671 & $-$5 & 0.09 &$2.15{\pm}0.22$   \\
NGC 4472 & 56    & 7453 & N4HD05 & 12:29:46.72 &    08:00:00.0 &  384 & 0.096 &  868 & $-$5 & 0.19 &$1.52{\pm}0.14$   \\
NGC 4527 &(56)& 7331 & N3ZD80 & 12:34:08.78 &    02:39:08.8 &  256 & 0.095 & 1734 &   +4 & 0.66 &\nodata   \\
NGC 4536 &(56)& 7331 & N3ZD0V & 12:34:27.10 &    02:11:16.9 &  384 & 0.079 & 1804 &   +4 & 0.57 &\nodata   \\
NGC 4565 & 235   & 7331 & N3ZD0W & 12:36:20.62 &    25:59:14.5 &  384 & 0.067 & 1227 &   +3 & 0.87 &\nodata   \\
NGC 4589 & 98    & 7886 & N4RW55 & 12:37:25.12 &    74:11:30.4 &  640 & 0.121 & 1980 & $-$5 & 0.19 &\nodata   \\
NGC 4594 &\nodata& 7331 & N3ZD86 & 12:39:59.40 & $-$11:37:21.0 &  384 & 0.223 & 1091 &   +1 & 0.59 &\nodata   \\
NGC 4636 & 152   & 7886 & N4RW58 & 12:42:49.83 &    02:41:14.6 &  640 & 0.124 & 1095 & $-$5 & 0.22 &\nodata   \\
NGC 4709 & 59    & 7453 & N4HD15 & 12:50:03.66 & $-$41:22:56.8 & 1600 & 0.512 & 4624 & $-$5 & 0.15 &\nodata   \\
NGC 4725 &(235)& 7330 & N3ZB98 & 12:50:26.79 &    25:30:05.4 &  640 & 0.051 & 1206 &   +2 & 0.29 &\nodata   \\
NGC 5193 &\nodata& 7453 & N4HD11 & 13:31:53.26 & $-$33:14:05.2 & 1920 & 0.242 & 3644 & $-$5 & 0.07 &\nodata   \\
IC 4296  & 225   & 7453 & N4HD12 & 13:36:38.86 & $-$33:57:55.9 & 1920 & 0.265 & 3871 & $-$5 & 0.05 &\nodata   \\
NGC 5273 &\nodata& 7330 & N3ZB16 & 13:42:08.53 &    35:39:13.9 &  640 & 0.044 & 1054 & $-$2 & 0.09 &\nodata   \\
NGC 5845 & 70    & 7886 & N4RW67 & 15:06:00.78 &    01:38:00.6 &  640 & 0.233 & 1450 & $-$5 & 0.35 &\nodata   \\
NGC 7014 & 82    &7453 & N4HD13 & 21:07:52.04 & $-$47:10:44.3 & 1600 & 0.142 & 4980 & $-$4 & 0.17 &\nodata   \\
NGC 7280 &\nodata& 7331 & N3ZD97 & 22:26:27.53 &    16:08:55.6 &  384 & 0.240 & 1844 & $-$1 & 0.31 &\nodata   \\
NGC 7331 &\nodata& 7450 & N41VB8 & 22:37:04.24 &    34:24:56.0 &  256 & 0.395 &  821 &   +3 & 0.65 &\nodata   \\
NGC 7457 &\nodata& 7450 & N41VB9 & 23:00:59.91 &    30:08:41.3 &  256 & 0.229 &  824 & $-$3 & 0.46 &\nodata   \\
NGC 7743 &\nodata& 7330 & N3ZB49 & 23:44:21.68 &    09:56:03.6 &  640 & 0.296 & 1662 & $-$1 & 0.15 &\nodata   \\

\enddata

\tablenotetext{a}{Groups and clusters as defined by Faber et al. 1989 
(Fornax=31, LeoI=57, Virgo=56); entries in parantheses are spirals not
classified by Faber et al. but clearly belonging to a group or cluster.}
\tablenotetext{b}{$B$-band extinction from Schlegel et al. 1998}
\tablenotetext{c}{T-types and ellipticities from RC3; ellipticity 
$e=(1{-}b/a)=(1{-}10^{{-}\log{R_{25}}})$}
\tablenotetext{d}{Kuntschner 2000, Kuntschner et al. 2000, and Trager et al. 2000}

\end{deluxetable}

\clearpage

\begin{deluxetable}{lcccccl}
\tablecolumns{7}
\tablewidth{0pc}
\tablecaption{Distances and F160W SBF Magnitudes\label{mbartable}}
\tablehead{
\colhead{Galaxy} &
\colhead{\vminio} &
\colhead{\mM} &
\colhead{\mbar} &
\colhead{\Mbar} &
\colhead{Dust} &
\colhead{Ref\tablenotemark{a}} \\
&
\colhead{(mag)} &
\colhead{(mag)} & 
\colhead{(mag)} &
\colhead{(mag)} & &
}
\startdata
NGC 7814 & $1.245{\pm}0.017$ & $30.44{\pm}0.14$ & $26.57{\pm}0.09$ & $-3.87{\pm}0.17$ &       D &$I$-SBF \\
NGC 221  & $1.133{\pm}0.007$ & $24.39{\pm}0.08$ & $19.11{\pm}0.05$ & $-5.28{\pm}0.10$ & \nodata &$I$-SBF \\
NGC 224  & $1.231{\pm}0.007$ & $24.24{\pm}0.08$ & $19.78{\pm}0.04$ & $-4.46{\pm}0.09$ & \nodata &$I$-SBF \\
         &                   & $24.38{\pm}0.05$ & $19.78{\pm}0.04$ & $-4.60{\pm}0.06$ &         &new PL \\
         &                   & $24.48{\pm}0.05$ & $19.78{\pm}0.04$ & $-4.70{\pm}0.06$ &         &new PL+Z \\
NGC 404  & $1.054{\pm}0.011$ & $27.41{\pm}0.10$ & $21.98{\pm}0.09$ & $-5.43{\pm}0.13$ & \nodata &$I$-SBF \\
NGC 524  & $1.221{\pm}0.010$ & $31.74{\pm}0.20$ & $27.19{\pm}0.20$ & $-4.55{\pm}0.28$ & \nodata &$I$-SBF \\
NGC 708  & $1.275{\pm}0.015$ & $33.83{\pm}0.20$ & $29.09{\pm}0.08$ & $-4.74{\pm}0.22$ &       D & HST $I$-SBF \\
NGC 821  & $1.196{\pm}0.022$ & $31.75{\pm}0.17$ & $27.03{\pm}0.09$ & $-4.72{\pm}0.19$ & \nodata &$I$-SBF \\
NGC 1052 & $1.213{\pm}0.010$ & $31.28{\pm}0.27$ & $26.51{\pm}0.07$ & $-4.77{\pm}0.28$ & \nodata &$I$-SBF \\
NGC 1172 & $1.112{\pm}0.032$ & $31.50{\pm}0.20$ & $26.57{\pm}0.09$ & $-4.93{\pm}0.22$ & \nodata &$I$-SBF \\
NGC 1316 & $1.132{\pm}0.016$ & $31.50{\pm}0.17$ & $26.11{\pm}0.09$ & $-5.39{\pm}0.19$ &       D &$I$-SBF \\
NGC 1351 & $1.148{\pm}0.016$ & $31.45{\pm}0.16$ & $26.68{\pm}0.05$ & $-4.77{\pm}0.17$ & \nodata &$I$-SBF \\
NGC 1339 & $1.134{\pm}0.012$ & $31.45{\pm}0.35$ & $26.43{\pm}0.07$ & $-5.02{\pm}0.36$ & \nodata &$I$-SBF \\
NGC 1344 & $1.135{\pm}0.011$ & $31.32{\pm}0.30$ & $26.26{\pm}0.07$ & $-5.06{\pm}0.31$ & \nodata &$I$-SBF \\
NGC 1365 & $1.153{\pm}0.028$ & $31.18{\pm}0.10$ & $25.64{\pm}0.09$ & $-5.54{\pm}0.10$ &       D & new PL \\
         &                   & $31.27{\pm}0.05$ & $25.64{\pm}0.09$ & $-5.63{\pm}0.10$ &         & new PL+Z \\
NGC 1373 & $1.085{\pm}0.013$ & $31.6\phn{\pm}0.5\phn$ & $26.40{\pm}0.12$ & $-5.2\phn{\pm}0.5\phn$ & \nodata &$I$-SBF \\
NGC 1374 & $1.146{\pm}0.016$ & $31.32{\pm}0.13$ & $26.50{\pm}0.13$ & $-4.82{\pm}0.18$ & \nodata &$I$-SBF \\
NGC 1375 & $1.070{\pm}0.019$ & $31.42{\pm}0.13$ & $25.95{\pm}0.07$ & $-5.47{\pm}0.15$ & \nodata &$I$-SBF \\
NGC 1379 & $1.143{\pm}0.019$ & $31.35{\pm}0.15$ & $26.24{\pm}0.11$ & $-5.11{\pm}0.19$ & \nodata &$I$-SBF \\
NGC 1380 & $1.197{\pm}0.019$ & $31.07{\pm}0.18$ & $26.43{\pm}0.05$ & $-4.64{\pm}0.19$ &       D &$I$-SBF \\
NGC 1380A& $1.138{\pm}0.018$ & $30.84{\pm}0.29$ & $26.16{\pm}0.16$ & $-4.68{\pm}0.33$ & \nodata &$I$-SBF \\
NGC 1381 & $1.189{\pm}0.018$ & $31.12{\pm}0.21$ & $26.55{\pm}0.10$ & $-4.57{\pm}0.23$ & \nodata &$I$-SBF \\
NGC 1386 & $1.101{\pm}0.018$ & $30.93{\pm}0.25$ & $25.81{\pm}0.06$ & $-5.12{\pm}0.26$ &       D &$I$-SBF \\
NGC 1387 & $1.208{\pm}0.047$ & $31.38{\pm}0.26$ & $26.0\phn{\pm}0.8\phn$ & $-5.4\phn{\pm}0.8\phn$ & D &$I$-SBF \\
NGC 1389 & $1.145{\pm}0.019$ & $31.52{\pm}0.18$ & $26.36{\pm}0.08$ & $-5.16{\pm}0.20$ & \nodata &$I$-SBF \\
NGC 1399 & $1.227{\pm}0.016$ & $31.34{\pm}0.16$ & $26.79{\pm}0.04$ & $-4.55{\pm}0.16$ & \nodata &$I$-SBF \\
NGC 1400 & $1.170{\pm}0.009$ & $31.95{\pm}0.33$ & $27.17{\pm}0.08$ & $-4.78{\pm}0.34$ & \nodata &$I$-SBF \\
NGC 1404 & $1.224{\pm}0.016$ & $31.45{\pm}0.19$ & $26.69{\pm}0.08$ & $-4.76{\pm}0.21$ & \nodata &$I$-SBF \\
NGC 1426 & $1.161{\pm}0.009$ & $31.75{\pm}0.18$ & $26.88{\pm}0.06$ & $-4.87{\pm}0.19$ & \nodata &$I$-SBF \\
NGC 1427 & $1.152{\pm}0.018$ & $31.70{\pm}0.24$ & $26.42{\pm}0.05$ & $-5.28{\pm}0.25$ & \nodata &$I$-SBF \\
IC 2006  & $1.183{\pm}0.018$ & $31.43{\pm}0.29$ & $26.61{\pm}0.05$ & $-4.82{\pm}0.29$ & \nodata &$I$-SBF \\
NGC 1553 & $1.159{\pm}0.016$ & $31.18{\pm}0.17$ & $26.32{\pm}0.07$ & $-4.86{\pm}0.18$ & \nodata &$I$-SBF \\
NGC 3031 & $1.187{\pm}0.011$ & $27.80{\pm}0.26$ & $22.99{\pm}0.05$ & $-4.81{\pm}0.26$ & \nodata & $I$-SBF \\
         &                   & $27.75{\pm}0.08$ & $22.99{\pm}0.05$ & $-4.76{\pm}0.09$ &         & new PL \\
         &                   & $27.80{\pm}0.08$ & $22.99{\pm}0.05$ & $-4.81{\pm}0.09$ &         & new PL+Z \\
NGC 3032 & $1.073{\pm}0.019$ & $31.55{\pm}0.28$ & $26.05{\pm}0.10$ & $-5.50{\pm}0.30$ &       D &$I$-SBF \\
NGC 3056 & $1.073{\pm}0.023$ & $30.27{\pm}0.25$ & $25.48{\pm}0.07$ & $-4.79{\pm}0.26$ & \nodata &$I$-SBF \\
NGC 3351 & $1.225{\pm}0.014$ & $29.85{\pm}0.09$ & $25.19{\pm}0.07$ & $-4.66{\pm}0.11$ &       D & new PL \\
         &                   & $30.00{\pm}0.09$ & $25.19{\pm}0.07$ & $-4.81{\pm}0.11$ &         & new PL+Z \\
NGC 3368 & $1.145{\pm}0.015$ & $29.92{\pm}0.22$ & $24.88{\pm}0.09$ & $-5.04{\pm}0.24$ &       D & $I$-SBF \\
         &                   & $29.97{\pm}0.06$ & $24.88{\pm}0.09$ & $-5.09{\pm}0.11$ &         & new PL \\
         &                   & $30.11{\pm}0.06$ & $24.88{\pm}0.09$ & $-5.23{\pm}0.11$ &         & new PL+Z \\
NGC 3379 & $1.193{\pm}0.015$ & $29.96{\pm}0.11$ & $25.26{\pm}0.08$ & $-4.70{\pm}0.14$ & \nodata &$I$-SBF \\
NGC 3384 & $1.151{\pm}0.018$ & $30.16{\pm}0.14$ & $25.34{\pm}0.17$ & $-4.82{\pm}0.22$ & \nodata &$I$-SBF \\
NGC 3928 & $1.096{\pm}0.015$ & $31.0\phn{\pm}0.6\phn$ & $25.60{\pm}0.09$ & $-5.4\phn{\pm}0.7\phn$ & D &$I$-SBF \\
NGC 4143 & $1.181{\pm}0.015$ & $30.85{\pm}0.19$ & $26.68{\pm}0.06$ & $-4.17{\pm}0.20$ & \nodata &$I$-SBF \\
NGC 4150 & $1.071{\pm}0.017$ & $30.53{\pm}0.24$ & $25.60{\pm}0.13$ & $-4.93{\pm}0.27$ &       D &$I$-SBF \\
NGC 4261 & $1.258{\pm}0.014$ & $32.34{\pm}0.19$ & $28.04{\pm}0.11$ & $-4.30{\pm}0.22$ & \nodata &$I$-SBF \\
NGC 4291 & $1.175{\pm}0.017$ & $31.93{\pm}0.32$ & $27.66{\pm}0.08$ & $-4.27{\pm}0.33$ & \nodata &$I$-SBF \\
NGC 4278 & $1.161{\pm}0.012$ & $30.87{\pm}0.20$ & $26.38{\pm}0.09$ & $-4.49{\pm}0.22$ & \nodata &$I$-SBF \\
NGC 4406 & $1.167{\pm}0.008$ & $31.01{\pm}0.14$ & $26.26{\pm}0.06$ & $-4.75{\pm}0.15$ & \nodata &$I$-SBF \\
NGC 4434 & $1.125{\pm}0.015$ & $31.98{\pm}0.17$ & $26.78{\pm}0.12$ & $-5.20{\pm}0.21$ & \nodata &$I$-SBF \\
NGC 4458 & $1.140{\pm}0.011$ & $31.02{\pm}0.12$ & $26.03{\pm}0.05$ & $-4.99{\pm}0.13$ & \nodata &$I$-SBF \\
NGC 4472 & $1.218{\pm}0.011$ & $30.90{\pm}0.10$ & $26.26{\pm}0.04$ & $-4.64{\pm}0.11$ & \nodata &$I$-SBF \\
NGC 4527 & $1.23\phn{\pm}0.03\phn$ & $30.53{\pm}0.09$ & $25.43{\pm}0.07$ & $-5.10{\pm}0.11$ & D & new PL/G\&S\tablenotemark{b} \\
         &                         & $30.61{\pm}0.09$ & $25.43{\pm}0.07$ & $-5.13{\pm}0.11$ &   & new PL/G\&S+Z \\
NGC 4536 & $1.20\phn{\pm}0.07\phn$ & $30.80{\pm}0.04$ & $25.46{\pm}0.12$ & $-5.34{\pm}0.13$ & D & new PL \\
         &                         & $30.87{\pm}0.04$ & $25.46{\pm}0.12$ & $-5.41{\pm}0.13$ &   & new PL+Z \\
NGC 4565 & $1.128{\pm}0.027$ & $31.05{\pm}0.17$ & $25.53{\pm}0.08$ & $-5.52{\pm}0.19$ &       D &$I$-SBF \\
NGC 4589 & $1.180{\pm}0.015$ & $31.55{\pm}0.22$ & $27.21{\pm}0.08$ & $-4.34{\pm}0.23$ & \nodata &$I$-SBF \\
NGC 4594 & $1.175{\pm}0.031$ & $29.79{\pm}0.18$ & $25.06{\pm}0.09$ & $-4.73{\pm}0.20$ & \nodata &$I$-SBF \\
NGC 4636 & $1.233{\pm}0.012$ & $30.67{\pm}0.13$ & $26.10{\pm}0.08$ & $-4.57{\pm}0.15$ & \nodata &$I$-SBF \\
NGC 4709 & $1.221{\pm}0.015$ & $32.88{\pm}0.17$ & $28.51{\pm}0.07$ & $-4.37{\pm}0.18$ & \nodata & HST $I$-SBF \\
NGC 4725 & $1.209{\pm}0.023$ & $30.45{\pm}0.34$ & $25.72{\pm}0.10$ & $-4.73{\pm}0.35$ & \nodata & $I$-SBF \\
         &                   & $30.38{\pm}0.06$ & $25.72{\pm}0.10$ & $-4.66{\pm}0.12$ &         & new PL \\
         &                   & $30.46{\pm}0.06$ & $25.72{\pm}0.10$ & $-4.74{\pm}0.12$ &         & new PL+Z \\
NGC 5193 & $1.208{\pm}0.015$ & $33.35{\pm}0.15$ & $28.52{\pm}0.06$ & $-4.83{\pm}0.16$ &       D & HST $I$-SBF \\
IC 4296  & $1.199{\pm}0.015$ & $33.53{\pm}0.16$ & $28.82{\pm}0.08$ & $-4.71{\pm}0.18$ &       D & HST $I$-SBF \\
NGC 5273 & $1.142{\pm}0.017$ & $30.93{\pm}0.26$ & $26.13{\pm}0.09$ & $-4.80{\pm}0.28$ & \nodata &$I$-SBF \\
NGC 5845 & $1.124{\pm}0.012$ & $31.91{\pm}0.21$ & $27.43{\pm}0.07$ & $-4.48{\pm}0.22$ & \nodata &$I$-SBF \\
NGC 7014 & $1.248{\pm}0.015$ & $33.84{\pm}0.15$ & $29.06{\pm}0.11$ & $-4.78{\pm}0.19$ & \nodata & HST $I$-SBF \\
NGC 7280 & $1.105{\pm}0.009$ & $31.77{\pm}0.22$ & $26.42{\pm}0.09$ & $-5.35{\pm}0.24$ & \nodata &$I$-SBF \\
NGC 7331 & $1.120{\pm}0.017$ & $30.43{\pm}0.17$ & $25.43{\pm}0.08$ & $-5.00{\pm}0.19$ &       D &$I$-SBF \\
         &                   & $30.81{\pm}0.09$ & $25.43{\pm}0.08$ & $-5.38{\pm}0.12$ &         & new PL \\
         &                   & $30.84{\pm}0.09$ & $25.43{\pm}0.08$ & $-5.41{\pm}0.12$ &         & new PL+Z \\
NGC 7457 & $1.104{\pm}0.009$ & $30.45{\pm}0.21$ & $25.57{\pm}0.09$ & $-4.88{\pm}0.23$ & \nodata &$I$-SBF \\
NGC 7743 & $1.080{\pm}0.009$ & $31.42{\pm}0.17$ & $25.99{\pm}0.06$ & $-5.43{\pm}0.18$ & \nodata &$I$-SBF \\

\enddata

\tablenotetext{a}{$I$-SBF=Tonry et al. 2001, shifted by $-$0.16 mag to the 
calibration derived using the Udalski et al. 1999 Cepheid period--luminosity relation;
HST $I$-SBF=Jensen et al. 2001 and Lauer et al. 1998, also calibrated using the new
period--luminosity relation; new PL=Freedman et al. 2001; 
new PL/G\&S=Gibson \& Stetson 2001; +Z indicates distances derived using 
a metallicity correction of $-0.2$ mag~dex$^{-1}$.}
\tablenotetext{b}{
The Cepheid period--luminosity relation and metallicity corrections used by Gibson \&
Stetson were the same as those used by Freedman et al. to determine the ``new PL''
and ``new PL+Z'' distances.  The Gibson \& Stetson distances assume a distance modulus for
the LMC of 18.45 rather than the value of 18.50 adopted by Freedman et al. 2001.
We have therefore added 0.05 mag to the Gibson \& Stetson values.
}
\end{deluxetable}

\end{document}